# Coupled models for total stress dissipation tests

Emoke Imre[1], Pál Rózsa[2+]

[1] HBM Systems Research Center, Obuda University, Hungary
[1] BME, Budapest, Hungary

**ABSTRACT:** Two linear, point-symmetric, coupled consolidation model families with various embedding space dimension values (oedometer models: 1, spherical models:3, cylindrical models:2), differing in one boundary condition (coupled 1: constant displacement, coupled 2: constant stress) are analysed analytically and numerically. The method of the research is partly analytical, the models are unified into a single model with unique analytical solution, every model can be derived from this by inserting the proper boundary condition and embedding space dimension *m*. The constants of the solutions are determined and an approximate time factor and model law are derived for the $m$ >1case which is identical to the one valid in the oedometer ($m$ =1) case. The convergence of the infinite series are examined in the function of the initial condition. Concerning the total stress at the pile shaft, significant decrease (with the value of the initial mean pore water pressure) is encountered for the coupled 1 consolidation models, zero stress drop is resulted by the coupled 2 models. The total stress dissipation test is suggested to be evaluated by the coupled 1 models with a time dependent constitutive law, eg., by adding a relaxation part-model. The rate of convergence is the smaller if the initial condition is the closer to the one of a zero solution (beyond the trivial one, a non-trivial zero solution exists for the coupled 1 model, at the Terzaghi‘s initial condition).

## 1. Introduction

### 1.1. Dissipation tests

The dissipation type tests are used for the laboratory/in situ assessment of permeability/coefficient of consolidation by evaluating the measured displacement or pore water pressure or total stress data with a consolidation model (Tables 1 to 5, Fig. 1, [1 to 11]).

Two kinds of staged oedometer tests are known with total stress load or with displacement load. In the conventional compression test, the total stress load is increased stepwise, the pore water pressure at the bottom, the displacement at the top of the sample are measured. In the oedometric relaxation test, the displacement load is increased stepwise, the pore water pressure at the bottom, the total stress at the top are measured.

The dissipation tests are made by stopping the steady penetration, clamping the cone penetrometer CPT system and measuring some stress variables in the function of the time. The stress variables are the local side friction $f_s$ and the cone resistance $q_c$, the pore water pressure u (CPTu), the total stress $\sigma$ and the pore water pressure u (the piezo-lateral stress cell (PSL) test) andthe total stress $\sigma$ and the pore water pressure u (the piezo-lateral stress cell at the flat dilatometer (DMT).

In the pore water pressure dissipation tests, the record is monotonic or non-monotonic with time, which is generally associated to low or high values of OCR, respectively. The u- sensors can be mounted in various positions, the corresponding dissipation curves are significantly different (eg., u1, u2, u3).

In the piezo-lateral stress cell test ([16]) the time variation of the radial total normal stress and the pore water pressure are measured. In the DMT dissipation test the time variation of the radial total normal stress is recorded. In soft clay, the radial total stress may decrease by 73%, the effective stress may vary non-monotonically ([11]), decreasing or increasing during the first few minutes, depending on the soil plasticity and OCR, the long term behaviour is not properly known. In other soils, very few pieces of information are available, the total stress may initially increase or the dissipation curve may not have inflextion point.

In the “simple rheological test” the time variation of the local side friction and the cone resistance are measured, the rod is clamped. Again, very few pieces of information are available. According to the results in relation to 2-minute long records, after an immediate stress drop, the cone resistance decreases, the shaft resistance decreases or increases in sand in the first minutes [19].

### 1.2. Dissipation test evaluation

#### 1.2.1. Models

The concept of linear, coupled 1 and coupled 2 consolidation model families is introduced for the folloowing displacement domains. The displacement domain of the point-symmetric consolidation models is bounded by spheres in 1,2,3 dimensions (Tables 3 to 5, Fig. 1). Constant displacement boundary condition is assumed at the inner boundary, which is at $r=r_0$ =0 for the two kinds of oedometer tests, being the symmetry point of a double-drained oedometric samples. The $r=r_0$ boundary is the surface of the model pile. Total stress load (coupled 2 models) or displacement load (coupled 1 models) is assumed at the outer boundary $r=r_1$, where in addition zero pore water pressure boundary condition is assumed. The pore water pressure solution of the coupled 2 models "reduces" to the one of the uncoupled model ([20]), so the uncoupled model family do not be needed to be discussed separately.

### 1.2.2. Evaluation

The analytical solutions of the oedometer case yield a dimensionless time variable $T^{oed}$ equations:

$$T^{oed} = \frac{ct}{H^2} \quad (1)$$

where $c$ is coefficient of consolidation, $H$ is model constant (Fig. 1). The non-linear parameter identification problem is solved approximately, in the lack of an automatic model. The one-point model fitting in practice requires time of $t_{90.}$

The analytical solution of the cylindrical and spherical cases do not yield similar time factor. The CPTu pore water pressure dissipation test are evaluated at present approximately with uncoupled models, using embedded initial conditions (generally assuming undrained penetration) and approximate time factors, two one-point fittings look like ([10, 15]):

$$c = \frac{T_{50}}{t_{50}} r_0^2 \quad (2)$$

$$c = \frac{T_{50}^{T-H}}{t_{50}} r_0^2 \quad I_r^{1/2} \quad (3)$$

where $r_0$ is the radius of the CPT equipment, $T_{50}$ is a time factor, and $t_{50}$ is the measured time for 50% dissipation, $I_r$ is the rigidity index.The time factors are heuristic, they are based on the observation so that the theoretical dissipation curves could be normalized. These include a model law for time t variable only.

The approximate, one-point model fitting requires time of $t_{50,}$ cannot handle if $t_{50}$< 50 s (partly drained penetration), and if the dissipation is starting from less than u0 values (the initial condition can not be varied as needed). Further problem is that it is difficult to assign a value of Ir since the shear modulus decreases with strains by a factor of 20 or 30 (Mayne, 2007).

The DMT total stress dissipation test evaluation method is model-free, it is based on an empirical formula concerning the inflexion point of the dissipation curve (and is not working in the lack of inflexion point).

### 1.2.3. Model validations

An automatic global minimisation algorithm was given to the injective solutions of linear PDE-s, giving reliability information as well [26].The non-linear parameter identification problem is solved mathematically precisely. The models of the two kinds of staged oedometer tests were validated against short multistage data with the results, that the linear models are acceptable for the pore water pressure, but for the total stress or displacement the relaxation or creep have to be taken into account ([22 to 26]).

The validation of the two kinds of cylindrical/spherical models against CPTu pore water pressure dissipation test data ended with the statement that both models are usable but the identified psrsmeters differ in a constant multiplier ([12, 13]).

The mean pore water pressure solution of the coupled 1 cylindrical model was used in an approximate way for the evaluation of dilatometer total stress dissipation test data with no inflexion point [18]. The total stress solution of the of the coupled 1 cylindrical model has not been used for the evaluation of the DMT or CPT total stress dissipation test data.

## 1.3. The aim and content of the paper

The total stress solution of the coupled 1 model has not been used for the evaluation of the total stress dissipation test. The cylindrical coupled model of Randolph-Wroth – the cylindrical analogon of the coupled Biot model for the oedometer compression test – gives constant total stress solution at $r=r_0$.

The analytical solution of the cylindrical and spherical coupled models do not yield time factor since the Bessel function roots are just nearly "regular". The „suggested, not derived"time factors are used with rigidity index and $r_0$ instead of the measure of the displacement domain. These include a model law for time t variable only.

Two kinds of coupled consolidation models are related to the staged oedometer tests differing in one boundary condition (coupled 1 – kinematic load - and 2 – total stress load). The hypothesis of the research that one of this which gives total stress dissipation, may qualitatively be good for the CPT dissipation tests. It is also assumed that a „more precise" time factor can be derived from the analytical solution, using the asymptotic Bessel formulae.

The aim of the paper is to analyze the two linear, pointsymmetric, coupled, linear model-families in terms of the initial condition, and displacement domain (undrained and partly drained cases) including the analytical and numerical properties. The solution is computed at embedding space dimension $m$=2 and compared with the case of other space dimensions using a suggested time factor.

The poperties of the analytical solution are determined in the function of the initial condition function qualitatively and quantitatively.

In this work it is shown that the cylindrical coupled 1 model can be used for the modelling of the dissipation around piles at $r=r_0$. The Terzaghi's time factor concept is extended to the cylindrical and, spherical case in a precise way.

The analytical solutions have basically the same numerical (convergence) properties within a model family. It is found that due to the similarity, the CPTu pore water dissipation test can even be evaluated by the oedometer model.

A unified mathematical formulation is given for the two coupled model families (i.e. two model sets with fixed boundary conditions and various space dimension m values).

In the first part of the paper the model analysis is given in the form of system of differential equations and analytical solution in terms of the dimension *m*. The structure of solution is treated. The analytical properties of the solution are qualitatively analysed for the two kinds of boundary conditions, independently of the embedding space dimension *m*.

In the second part of the paper some simulations are made. The constants of the solutions are presented in the function of the boundary conditions for embedding space dimension *m*=2. Approximate closed form solutions for the boundary condition equations are given which are the same for the same boundary condition (within a model family), resulting a time factor *T*. The convergence properties are characterized.

The practical significance is that the analytical models can be used in the precise evaluation of the pore water pressure / total stress dissipation tests with the identification of the initial condition. In this way the evaluation methods can be used to reduce of the test duration. Concerning the evaluation of the total stress dissipation test, an example is showen with the coupled 1 models used for DMT data.

**Table 1** Types of one dimensional oedometric dissipation tests with constant boundary condition

| | |
|---|---|
| (Multistage) relaxation test | (MRT) |
| (Multistage) compression test | (MCT) |

**Table 2** Types dissipation tests made with static penetrometers, modelled with cylindrical and spherical (ellipsoid) shaped domain.

| Measured variable | dissipation test |
|---|---|
| Pore water pressure u, sensor on the shaft and/or on the tip | CPTu dissipation test |
| Total stress $\sigma$ and pore water pressure u, effective stress sensor on the shaft | piezo-lateral stress cell test |
| Total stress $\sigma$, sensor on the shaft | $\sigma$ DMT dissipation test after A or B position |
| the local side friction $f_s$ and the cone resistance $q_c$ | CPT $f_s$ (effective stress) and $q_c$ dissipation tests |

**Table 3** 1D point-symmetric consolidation models

| *V* or $\varepsilon$ boundary condition | 1D (Oedometric models) |
|---|---|
| no (uncoupled) | Terzaghi (1923 ) [4] |
| *v-v* (coupled 1) | Imre ( 1997-1999) [5] |
| *v*-$\varepsilon$ (coupled 2) | Biot (1941) [6] |

**Table 4** 2D point-symmetric consolidation models

| *V* or $\varepsilon$ boundary condition | 2D (Cylindrical pile models) |
|---|---|
| no (uncoupled) | Soderberg (1962) [7] |
| *v-v* (coupled 1) | Imre & Rózsa (1998) [2] |
| *v*-$\varepsilon$ (coupled 2) | Randolph at al (1979) [1,8] |

**Table 5** 3D consolidation models

| *v* or $\varepsilon$ boundary condition | 3D (Spherical pile models) |
|---|---|
| no (uncoupled) | Torstensson (1975) [9] |
| *v-v* (coupled 1) | Imre & Rózsa (2002) [3] |
| *v*-$\varepsilon$ (coupled 2) | Imre & Rózsa (2005) [10] |

**Table 6.** Cylindrical model.Initial condition series in terms of parameter D [-]

| $u_0$ | 1 | 3 | 4 | 7 | 8 | 9 | 10 |
|---|---|---|---|---|---|---|---|
| D | 0.001 | 0.053 | 0.135 | 0.408 | 0.591 | 740 | 0.970 |

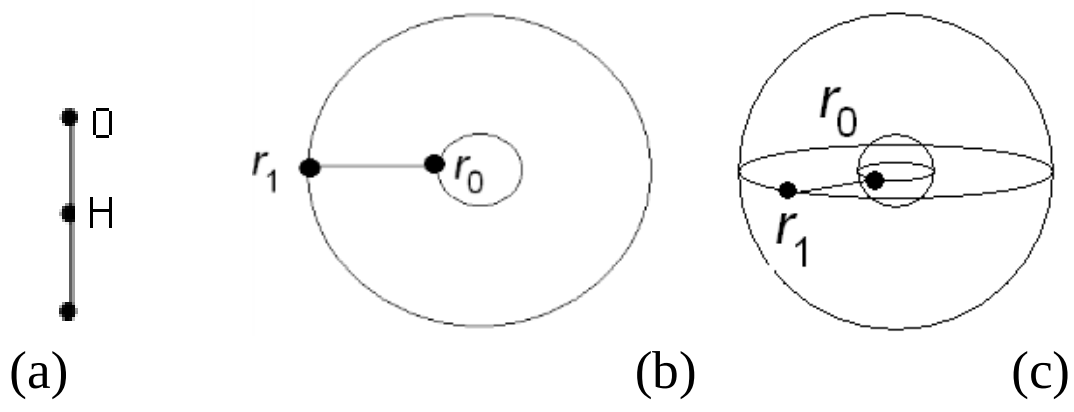

Fig. 1. The displacement domain bounded by a (a) 0 dimensional sphere (oedometer model), (b) 1 dimensional sphere (cylindrical model), (c) 2 dimensional sphere (spherical model).

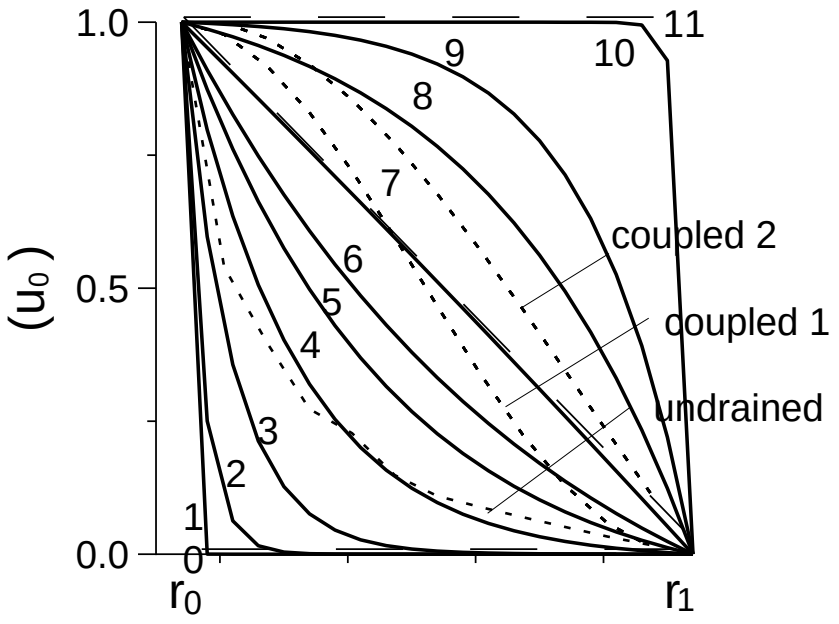

Fig. 2. Cylindrical model. Initial pore water pressure function series, with the ones of the one-term solutions and the one determined by undrained penetration modelling, *n*= $r_1$ / $r_0$=37, $r_0$=1.75 cm.

## 2. Unified formulation

### 2.1. System of differential equations

Two unified equations can be derived ([12]). Equation (1) compiles the equilibrium condition, the effective stress equality, the geometrical and, the constitutive equations, as follows:

$$E_{oed}\frac{\partial \varepsilon}{\partial r} - \frac{\partial u}{\partial r} = 0 \qquad (4)$$

and, Equation (4) compiles the continuity equation, the Darcy's law (neglecting the gravitational component of the hydraulic head) and the geometrical equation, as follows:

$$-\frac{k}{\gamma_v}\Delta\ u + \frac{\partial \varepsilon}{\partial t} = 0 \qquad (5)$$

where the volumetric strain and the Laplacian operator are dependent on *m* as follows:

$$\varepsilon = \frac{1}{r^{m-1}}\frac{\partial}{\partial r}\left(r^{m-1}v\right) \qquad (6)$$

$$\Delta = \frac{1}{r^{m-1}}\frac{\partial}{\partial r}\left(r^{m-1}\frac{\partial}{\partial r}\right) \qquad (7)$$

*v* is the radial displacement, *u* is the excess pore water pressure, *r* and *t* are the space and the time co-ordinates respectively, $E_{oed}$ is the oedometric modulus:

$$E_{oed} = \frac{2G(1-\mu)}{1-2\mu} = \frac{E(1-\mu)}{(1+\mu)(1-2\mu)}, \mu < 0.5 \qquad (8)$$

*G* is the shear modulus, *E* is the is Young modulus, $\mu$ is the Poisson's ratio.

## 2.2. Boundary conditions

Four boundary conditions are presented for $m$=2.
(1) The (common) *boundary condition Nr. 1* implies that the pore water pressure is zero at $r= r_1$:

$$u(t,r)|_{r=r_1} = 0 \tag{9}$$

(2) The (common) *boundary condition Nr. 2* entails that the flux is equal to zero at $r= r_0$ :

$$\frac{\partial u(t,r)}{\partial r}\Big|_{r=r_0} \equiv 0 \tag{10}$$

(3) The (common) *boundary condition Nr. 3* implies that the displacement equals to a constant at $r= r_0$:

$$v(t,r)|_{r=r_0} \equiv v_0 > 0 \tag{11}$$

*(4) Boundary condition Nr. 4* – concerning the Imre-Rózsa model – implies that the displacement is zero at $r= r_1$:

$$v(t,r)|_{r=r_1} \equiv 0 \tag{12}$$

(5) *Boundary condition Nr. 5* - concerning the Randolph-Wroth model – expresses that the volumetric strain $\varepsilon$ is constant $r= r_1$:

$$\varepsilon(t,r)|_{r=r_1} \equiv \varepsilon_1 > 0. \tag{13}$$

These boundary conditions are equally usable for $m$=3, in the case of $m$=1, $r_0$ = 0 is assumed, the half of the space domain is used.

## 2.3. Structure of Solution

The structure of the solution can be determined on the basis of the concept of linear ordinary differential equations ([5]). The solution is equal to the following sum for each variable:

$$\bullet(t,r) = \bullet^p(r) + \bullet^L(r) + \bullet^t(t,r) + \bullet^w \tag{14}$$

where the superscripts $p$ or $L$ indicate the steady-state, drained continuum-mechanical or seepage problems resp., $t$ indicates transient, $w$ concerns the self-weight component.

The solution of the transient part satisfies the homogeneous form of the boundary conditions and its final value is zero. The solution of the steady-state part satisfies the inhomogeneous form of the boundary conditions.

## 3. Qualitative features of Solution

## 3.1. Methods

The displacement solution is expressed in terms of the pore water pressure solution. The $m$=2 case is considered, however, nearly all equations are valid for every space dimension.

The initial condition for the pore water pressure is assumed to be given in the form of the following monotonic normalised parametric functions for both the qualitative and numerical analysis in this work, if it is not indicated otherwise:

$$u_0(r,F) = \frac{1 - e^{\frac{r-r_1}{F}}}{1 - e^{-\frac{r_1-r_0}{F}}} \qquad F \neq 0; \quad r_0 \leq r \leq r_1 \tag{15}$$

The value for $u_0$ at $r= r_0$ is equal to 1. The value for $u_0$ at $r= r_1$ is equal to 0. Analysing the shape function of the pore water pressure initial condition, it can be observed that parameters $F$ and $D$ are in one-to one relation.The mean normalised initial pore water pressure for embedding space dimension $m$=2:

$$D(F,r_0,r_1) = \frac{2}{r_1^2 - r_0^2}\int_{r_0}^{r_1} r u_0(r,F)\, dr \tag{16}$$

In the qualitative and quantitative analyse a normalised shape function series is used with the mean, normalised initial pore water pressure D series are shown in Table 6.

The monotonic initial condition series for the pore water pressure was given in the form of the monotonic, normalised, parametric functions (Eq 11).

Analysing the shape functions of the pore water pressure initial condition, it can be observed that parameters $F$ and $D$ are in one-to one relation for fixed space domains.

Ten values for the mean of the initial pore water pressure (parameter $D$) were selected, the value of parameter $F$ was determined for each value of $D$ and, for each value of $r_1$. The initial pore water pressure functions are shown in Figure 2 in the case of $n$=37.

The shape functions are strictly convex, or concave, or linear. At the limits $F\to$-0 and, $F\to$+0, it is constant $u_0(r)\equiv 0$ and, $u_0(r)\equiv 1$, respectively, it is linear at the limit where $F$ equals to the plus or minus infinity (the $D$ value is about equal to 0.33 and 0.5 in the cylindrical and oedometric case, resp.).

## 3.2. Results

### 3.2.1. Analysis of Equilibrium Equation

By integrating the modified equilibrium Equation (4) with respect to $r$ including boundary condition *Nr.* 1:

$$u(t,y) = E_{oed}\,\varepsilon(t,r) - E_{oed}\,\varepsilon(t,r)|_{r=r_1} \tag{17}$$

## Coupled 1 model

A boundary condition function is derived by further integration between $r_0$ and $r_1$ using boundary condition *Nr.* 3 and boundary condition *Nr.* 4:

$$\varepsilon(t,r)|_{r=r_1} = -\left(\frac{u_{mean}(t)}{E_{oed}}\right) \tag{18}$$

where:

$$\varepsilon(t,r)|_{r=r_1} = -\left(\frac{u_{mean}(t)}{E_{oed}}\right) \tag{19}$$

$$u_{mean}(t) = \int_{r_0}^{r_1} u(t,r)\, dr \Big/ \int_{r_0}^{r_1} r\, dr \tag{20}$$

Inserting this boundary condition function into the equilibrium Equation:

$$u(t,r) = E_{oed}\,\varepsilon(t,r) + u_{mean}(t) \quad (21)$$

From this:

$$\varepsilon(t,r) = \frac{1}{E_{oed}}\left[u(t,r) - u_{mean}(t)\right] \quad (22)$$

It follows that for a realistic $u$ the change in $\varepsilon$ with $t$ is positive in the vicinity at the outer boundary (rebound) and negative in the vicinity of the pile (compression). The final value of the transient strain is zero. By further integration:

$$v(t,r) = \frac{1}{rE_{oed}}\left(\int_{r1}^{r} ru(t,r)\,dr - u_{mean}(t)\int_{r1}^{r} r\,dr\right) \quad (23)$$

It follows from the analysis of Equation (1) that for a realistic $u$ the transient part of $v$ is non-negative and, monotonously decreases with $t$ for any $r$. The initial condition functions for $u$ and $v^t$ have the following relationships:

$$u_0(r) = E_{oed}\,\varepsilon_0^t(r) - E_{oed}\,\varepsilon_0^t(r)|_{r=r_1} \quad (24)$$

$$v_0^t(r) = \frac{1}{rE_{oed}}\left(\int_{r1}^{r} ru_0(r)\,dr - u_{0,mean}\int_{r1}^{r} r\,dr\right) \quad (25)$$

The Terzaghi's initial condition – where $u_0(r)$ is uniform with a positive value of $c_1$ – yields the following initial displacement function $v^t_0(r)$ for the coupled 1 model:

$$v_0^t(r) \equiv 0 \quad (26)$$

This is the zero solution function and the meaning is that the dissipation is instantaneous.

## Coupled 2 model

Inserting the inhomogeneous form of boundary condition *Nr.* 5 into the equilibrium Equation:

$$u(t,y) = E_{oed}\,\varepsilon(t,r) \quad (27)$$

From this:

$$\varepsilon(t,y) = \frac{u(t,r)}{E_{oed}}. \quad (28)$$

It follows from the analysis of the equilibrium Equation that for a realistic $u$ the change in $\varepsilon$ with $t$ is -negative (compression). By further integration:

$$v(t,r) = \frac{1}{rE_{oed}}\int_{r_0}^{r} ru(t,r)\,dr \quad (29)$$

It follows from the analysis of the equilibrium Equation that for a realistic $u$ the transient part of $v$ is non-negative and, monotonously decreases with $t$ for any $r$. The initial condition functions for $u$, $\varepsilon^t$ and $v^t$ have the following relationships:

$$u_0(r) = E_{oed}\,\varepsilon_0^t(r) \quad (30)$$

$$v_0^t(r) = -\frac{1}{rE_{oed}}\int_{r_0}^{r} ru_0(r)\,dr. \quad (31)$$

The Terzaghi's initial condition – where the initial pore water pressure function $u_0(r)$ is uniform with a positive value of $c_1$ – yields the following initial displacement function $v^t_0(r)$ for the coupled 2 model:

$$v_0^t(r) = -\frac{c_1 r}{2E_{oed}}. \quad (32)$$

For a monotonic, positive initial pore water pressure function $u_0(r)$, compression and, as a result, in the vicinity of $r_1$ inward displacement takes place, the volume of the displacement domain is decreasing. The final value of the transient strain is zero.

### 3.2.2. Analysis of Continuity Equation

The modified continuity Equation (5) can be written as follows by inserting the time dependent part of the volumetric strain $\varepsilon^t$:

$$-\frac{k}{\gamma_w}\Delta\, u + \frac{\partial \varepsilon^t}{\partial t} = 0 \quad (33)$$

By integrating the resulting equation twice with respect to $r$ using the homogenous form of the boundary conditions *Nr.* 2 and *Nr.* 1 respectively, the following explicit expression can be derived for the pore water pressure:

$$u(t,r) = -\frac{\gamma_w}{k}\int_{r_1}^{t}\frac{1}{k}\int_{r_0}^{k} x\frac{\partial \varepsilon^t(t,x)}{\partial t}\,dxdk. \quad (34)$$

The pore water pressure function is further integrated with respect to $t$ between 0 and $\infty$:

$$\int_0^{\infty} u(t,r)\,dt = -\frac{\gamma_w}{k}\int_{r_1}^{x}\frac{1}{k}\int_{r_0}^{k}[0 - x\varepsilon^t(0,x)]\,dxdk \quad (35)$$

The initial value for the transient volumetric strain (which can be written in terms of the initial pore water pressure) characterizes the rate of consolidation in every point. Especially, there is no time dependent consolidation if this function is the zero function.

Ratio of functional (33) for the initial condition series shown in Table 6 in case of the coupled 1/coupled 2 model is shown in Table 7. The result indicates that the rate of dissipation is increasingly larger for the coupled 2 model with increasing value of parameter $D$.

### 3.2.3. Total and effective stress solutions

The stress-state variable of the constitutive equation of the saturated soils is the effective stress $\sigma'$, the difference of the total stress and the pore water pressure $\sigma'=\sigma-u$ where $\sigma$ is the total stress. The compression strain is positive. On the basis of $u$ and $\varepsilon$ the solutions, the total stress $\sigma$ and the effective stress $\sigma'$ can be assessed using the effective stress equality and the constitutive equations for embedding space dimension $m$=2:

$$\sigma = \sigma' + u. \quad (36)$$

Table 7. Cylindrical model, dissipation test duration: ratio coupled 1/coupled 2 of functional (33) for the initial condition series in Table 6

| $u_0$ | 1 | 3 | 4 | 8 | 9 | 10 |
|---|---|---|---|---|---|---|
| D [-] | 0.001 | 0.053 | 0.135 | 0.591 | 0.740 | 0.970 |
| ratio Eq (33) | 0,85 | 0,66 | 0,55 | 0,28 | 0,20 | 0,03 |

$$\sigma'_r = -\frac{2G}{1-2\mu}\left[(1-\mu)\varepsilon - (1-2\mu)\frac{v}{r}\right] \quad (37)$$

$$\sigma'_{\varphi} = -\frac{2G}{1-2\mu}\left[\mu\varepsilon + (1-2\mu)\frac{v}{r}\right] \quad (38)$$

$$\sigma'_{z} = -\frac{2G\mu}{1-2\mu}\varepsilon \quad (39)$$

For the coupled 1 model, the effective stresses at the shaft-soil interface:

$$\sigma'^{t}_{r}(t,r)|_{r=r_0} = -[u(t,r_0) - u_{mean}(t)] \quad (40)$$

$$\sigma'^{t}_{\varphi}(t,r)|_{r=r_0} = -\frac{\mu[u(t,r_0) - u_{mean}(t)]}{1-\mu} \quad (41)$$

It follows that for a realistic $u$ the transient effective stress is negative around the shaft with zero final value and the effective stress at the shaft-soil interface increases with time here. It also follows that the effective stress at the outer boundary decreases with time, the mean of the first invariant of the effective stress tensor on the displacement domain is constant.

For the coupled 1 model, the radial total stress at the shaft-soil interface:

$$\sigma^{t}_{r}(t,r)|_{r=r_0} = u_{mean}(t) \quad (42)$$

It follows that for a realistic $u$ the radial total stress at the shaft-soil interface decreases with time.

For the coupled 2 model, the effective stresses at the shaft-soil interface:

$$\sigma'^{t}_{r}(t,r)|_{r=r_0} = -u(t,r) \quad (43)$$

$$\sigma'^{t}_{\varphi}(t,r)|_{r=r_0} = -\frac{\mu}{1-\mu}u(t,r) \quad (44)$$

It follows that for a realistic $u$ the transient effective stress is negative around the shaft with zero final value and the effective stress at the shaft-soil interface increases with time.

For the coupled 2 model, the radial total stress at the shaft-soil interface is constant with time.

$$\sigma^{t}_{r}(t,r)|_{r=r_0} = 0 \quad (45)$$

It follows that the radial total stress at the shaft-soil interface is constant.

## 4. Analytical Solution

### 4.1. Steady-state solution part

The solution of the drained continuum-mechanical problem for the displacement $v^p$ is the solution of the following part of Equation (1):

$$E_{oed}\frac{\partial\varepsilon}{\partial r} = 0 \quad (46)$$

which is the cavity expansion model for $m$=2, 3 and the oedometer ($K_0$) compression model for $m$=1. The solution has the following general form:

$$v^p = \frac{\alpha}{r^{m-1}} + \beta r \quad (47)$$

where the parameters can be determined from the inhomogeneous form of the boundary conditions (i.e. the common boundary condition *Nr.* 3 and, *Nr.* 4 - Imre-Rózsa model , *Nr.* 5 - Randolph-Wroth model).

These can be rewritten in the following form for $m$=2 as follows. The displacement $v^p$ for the Imre-Rózsa model:

$$v^p(r) = \frac{r_0 v_0}{r_1^2 - r_0^2}\left(\frac{r_1^2}{r} - r\right). \quad (48)$$

and, for the Randolph-Wroth model:

$$v^p(r) = \frac{r_0 v_0}{r} + \frac{\varepsilon_1 r_0^2}{2r} + \frac{\varepsilon_1}{2}r. \quad (49)$$

The solution of the steady-state seepage problem for the pore water pressure $u^L$ is identically equal to zero since the hydrodynamic boundary conditions are homogeneous. Therefore, superscript t is omitted for the pore water pressure in the following.

## 4.2. Transient solution part

### 4.2.1. Analytical solution

The transient part of the displacement solution in the function of $n$ (Imre et al, 2007):

$$v^t(t,r) = r^{\frac{-(m-2)}{2}} \sum_{k=1}^{\infty} C_k [J_{m/2}(\lambda_k r) + \mu_k Y_{m/2}(\lambda_k r)]\, e^{-[\lambda_k]^2 ct}$$

(50)where $J_{m/2}$ and $Y_{m/2}$ are the Bessel functions of the first and second kinds, with the order of $m/2$, and $\lambda_k$, $\mu_k$, $C_k$ parameters of the solution, $m$ is embedding space dimension, $c$ is coefficient of consolidation.

The volumetric strain and the pore water pressure solution from this:

$$\varepsilon^t(t,r) = r^{\frac{-(n-2)}{2}} \sum_{k=1}^{\infty} C_k \lambda_k \left\{[J_{(m-2)/2}(\lambda_k r) + \mu_k Y_{(m-2)/2}(\lambda_k r)]\right\} e^{-[\lambda_k]^2 ct} \quad (51)$$

The function $u$ is then determined using Equation (1). For the Imre-Rózsa (i.e. $m$=2, coupled 1) model:

$$u(t,r) = \sum_{k=0}^{\infty} \lambda_k C_k\, e^{-\gamma_k^2 \cdot c_h \cdot t} \left\{[I_0(\lambda_k r) + \mu_k Y_0(\lambda_k r)] - [I_0(\lambda_k r_1) + \mu_k Y_0(\lambda_k r_1)]\right\}$$

(52)and, where $c_h$ is coefficient of consolidation. For the Randolph-Wroth (i.e. $m$=2, coupled 2) model:

$$u(t,r) = \sum_{k=0}^{\infty} \gamma_k D_k\, e^{-\gamma_k^2 \cdot c_h \cdot t} [I_0(\gamma_k r) + \beta_k Y_0(\gamma_k r)]. \quad (53)$$

### 4.2.2. Constants of solutions

The parameters $C_k$ of the solution can be determined from the initial condition. The parameters $\lambda_k$, $\mu_k$ of the solution can be determined from the boundary conditions.

The parameters $C_k$ of the solution can be determined from the initial condition as follows. The initial displacement functions $v^t_0$ $(r)$ can be determined from $u_0(r)$ with the use of Eqs 21 and 27 . The coefficients $C_k$ can be determined using the orthogonality of the solution functions. In the case of $m$=2, the Bessel coefficients $C_k$ and $D_k$ from initial displacement function $v^t_0(r)$:

$$C_k orD_k = \frac{\int_{r0}^{r1} r\, v_0^t(,r)\left[J_1(\lambda_k r)+\mu_k Y_1(\lambda_k r)\right]dr}{\int_{r0}^{r1} r\left[J_1(\lambda_k r)+\mu_k Y_1(\lambda_k r)\right]^2 dr} \quad (54)$$

The parameters $\lambda_k$, $\mu_k$ of the solution can be determined from the boundary conditions as follows. For the coupled 1 or 2 model-families, the "boundary condition equation" (arisen from the homogeneous form of boundary conditions Nr.3 and Nr.4 or Nr.3 and Nr.5) can be written as follows, respectively:

$$J_{m/2}(\lambda_k r_0)Y_{m/2}(\lambda_k r_1) - J_{m/2}(\lambda_k r_1)Y_{m/2}(\lambda_k r_0) = 0 \quad (55)$$

$$J_{m/2}(\lambda_k r_0)Y_{(m-2)/2}(\lambda_k r_1) - J_{(m-2)/2}(\lambda_k r_1)Y_{m/2}(\lambda_k r_0) = 0 \quad (56)$$

The roots of the boundary condition equation for the coupled 1 and 2 model-families, respectively, for $m=1$:

$$\lambda_k = \frac{k\pi}{(r_1 - r_0)}, \quad \gamma_k = \frac{(2k-1)\pi}{2(r_1 - r_0)} \quad (57)$$

Approximate closed form solution can be suggested as follows for m>1. The asymptotical Bessel function formulae:

$$J_n(r) = \sqrt{\frac{2}{\pi r}}\cos\left(r - \frac{\pi}{4} - \frac{n\pi}{2}\right) \quad (58)$$

$$Y_n(r) = \sqrt{\frac{2}{\pi r}}\sin\left(r - \frac{\pi}{4} - \frac{n\pi}{2}\right) \quad (59)$$

Using the asymptotical Bessel function formulae, the approximate form for the BCE for the coupled 1 and 2 model-families, respectively, for the m>1 case:

$$\sin(\lambda_k (r_1 - r_0)) = 0 \quad (60)$$

$$\sin(\gamma_k (r_1 - r_0) + \pi/2) = 0 \quad (61)$$

being equally valid for dimensions embedding space dimension $m=2$ or 3. The roots for the coupled 1 and 2 model-families, respectively:

$$\lambda_k \approx \frac{k\pi}{(r_1 - r_0)}, \quad \gamma_k \approx \frac{(2k-1)\pi}{2(r_1 - r_0)} \quad (62)$$

Within a model-family, the 1 dimensional and the approximate 2 and 3 dimensional formulae are identical. Inserting into the analytic solutions of the coupled 1 models, some dimensionless variables are resulted:

$$(r) = \frac{r}{r_1 - r_0} \text{ and } (r) = \frac{r}{2(r_1 - r)_0} \quad (63)$$

$$T = \frac{ct}{(r_1 - r_0)^2} \text{. and } T = \frac{ct}{4(r_1 - r_0)^2} \quad (64)$$

These formulae reflect that the rate of dissipation is faster for the coupled 1 than for the coupled 2 models.

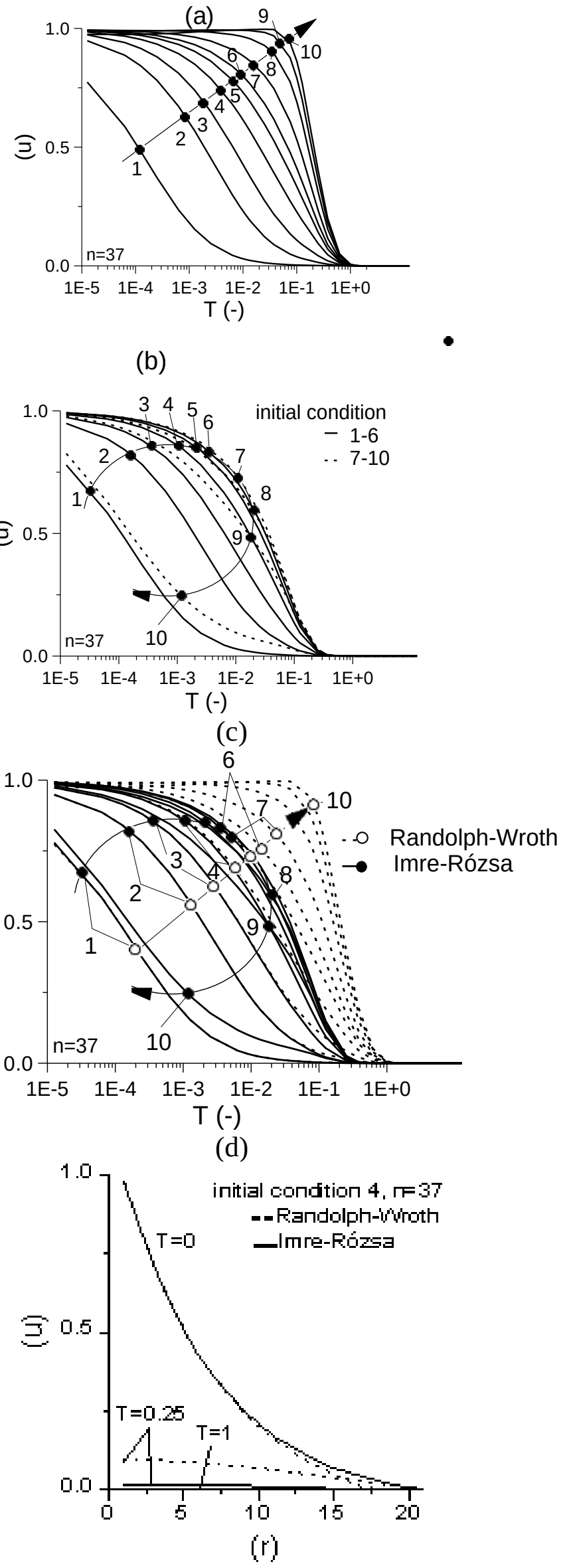

Fig. 3. Cylindrical pore water pressure solution. (a) to (c). Dissipation curves in the function of the initial condition for the Randolph-Wroth model, for the Imre-Rózsa model, comparing the results, resp. (d) Variation of the $u$ with $t$ and $r$.

(a)

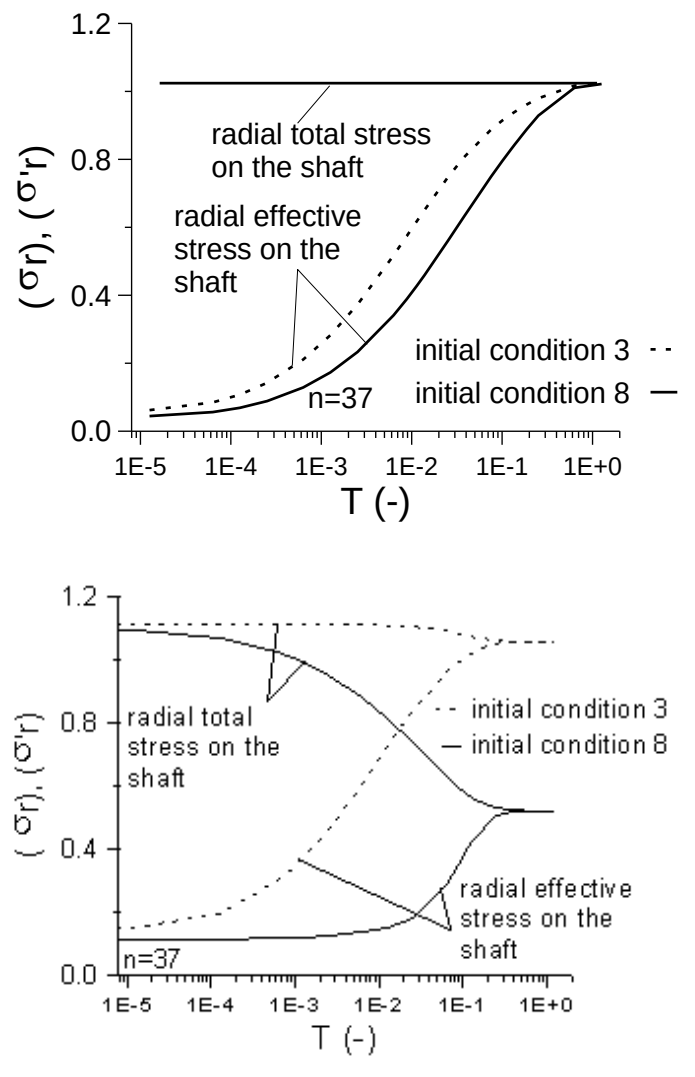

(b)

Fig. 4. (a) and (b) The time variation of the transient component of the radial normal stresses acting on the shaft, cylindrical a. Randolph-Wroth model. b. Imre-Rózsa model, initial condition 8.

## 5. Simulations

### 5.1. Simulation methods

The space domain is defined with radius $r_0$ =1.75 and $r_1$ = 64.75 in the case of undrained penetration, well above the tip. Some numerical examples are given to illustrate the solution for $m$ =2, coupled 1 (Imre-Rózsa) model and coupled 2 (Randolph-Wroth) model, in the function of the monotonic inital initial pore water pressure function series, since the initial condition may vary in the function of the soil properties even in case of undrained penetration. Similar analyses were made for the oedometric case in ([5]).

The monotonic initial condition series for the pore water pressure was given in the form of the monotonic, normalised, parametric functions (Eq 11). The mean ordinate was used to specify the numerical examples (Eq 12, Table 6).

The sum of the solution of the drained continuum-mechanical problem and the self-weight was set to be equal to the initial value of the piezo-lateral stress cell measurement (see Fig 16, [15]). A parametric analysis was made, the model constant displacement $v_0$ was computed for both models assuming $\mu$=0.3, $G$=50 kPa and various values for $K_0$.

The following normalized space coordinate was applied representing the results on the space domain:

$$(r) = \frac{20(r - r_0) + (r_1 - r_0)}{r_1 - r_0} \qquad (65)$$

### 5.2. Simulation results

The results are presented for $m$=2, on the example of $r_0$ =1.75 cm and $r_1$ = 64.75 cm, $n = r_1 / r_0 = 37$ (this space domain is related to the undrained penetration problem of the CPT, well above the tip [21].

### 5.2.1. Pore water pressure at the shaft

The initial pore water pressure functions are shown in Figure 2 where the initial conditions of the one-term solution moreover, the initial pore water pressure distribution after undrained penetration determined by the strain path method are also indicated.

According to the results, the one-term solutions and the pore water pressure distribution after undrained penetration determined by the strain path method are roughly found in the strip of the four not too extreme initial conditions (i.e. 4 to 7, see Fig 2). Therefore, the one-term solution, the solution related to the initial condition of the undrained penetration and, the solutions related to the initial conditions 4 to 7 are similar to each-other and can be interchanged.

According to the results of the qualitative analysis, the rate of the pore water pressure dissipation on the shaft is controlled by the initial, transient mean effective stress, which depends differently on the initial condition for the two models. The pore water pressure dissipation functions are shown in Figure 3.

According to the results of the simulation, the pore water pressure dissipation curves are in accordance to this, the dissipation is faster for the coupled 1 model than the coupled 2 model at fixed initial condition.

As the distance from the zero solution is increasing (ie., with increasing $D$), the dissipation time in terms of time factor ($T_\kappa$) increases for any fixed $\kappa$ (i.e. the curves “move” from left to right, see Fig 3).

For the coupled 1 model, having two zero solutions, the linear initial condition separates the dissipation curves solutions. If $D$<0.33 (i.e. convex initial distributions) or if $D$<0.33 (i.e. concave distributions) the dissipation time increases/decreases with increasing $D$. As a result, the dissipation curves related to some concave and convex initial condition functions coincide (Fig 4(b)).

The common features of the solution of the two models are as follows. According to the results shown in Figure 3 (a) to (c), for the not too extreme initial conditions (i.e. 4 to 7), the dissipation curve solutions are very similar, they nearly coincide. Especially, the dissipation curves coincide at great degrees of dissipation ($\kappa\approx$99,9%, $\kappa=(u_{max}-u)/u_{max}$). The time factor $T_\kappa$ is about three times larger for the Randolph-Wroth’s model, than for the Imre-Rózsa model resulting in larger dissipation times. It follows that both models can be used for the evaluation of the the pore water pressure dissipation tetsts, the identified parameters will slightly differ.

### 5.2.2. Total stress at the shaft

According to the results of the qualitative analysis, the rate of total stress dissipation on the shaft is controlled by the initial pore water pressure distribution differently for the two models.

Concerning the transient part of the solution for total stress, in the case of the Randolph-Wroth model it is the zero function and, as a result, the radial total normal stress at $r_0$ is constant and, therefore, the radial effective stress at $r_0$ increases by the value of the initial pore water pressure at $r_0$. These features are unrealistic for soft clays (Fig. 4a).

Concerning the transient part of the solution for total stress, in the case of the Imre-Rózsa model, it is the mean pore water pressure. As a result, the radial total normal stress at $r_0$ decreases with time by the value of initial mean pore water pressure depending linearly on $D$ and, the radial effective stress at $r_0$ increases with time by the value of the difference of the initial mean pore water pressure and the initial pore water pressure at $r_0$ (Fig. 4b).

### 5.2.3. Stresses within the soil

The solutions for the transient displacement $v^t$ and volumetric strain $\varepsilon^t$ are shown in Figure 5 and 6. The initial value of the transient part of the radial displacement $v^t$ is non-negative, the final value is identically equal to zero. Therefore, the transient displacement $v^t$ basically decreases with $t$ for any $r$ up to zero in the case of both models. As a result, the displacement at $r_1$ is decreasing for the Randolph–Wroth model (inward moving outer boundary).

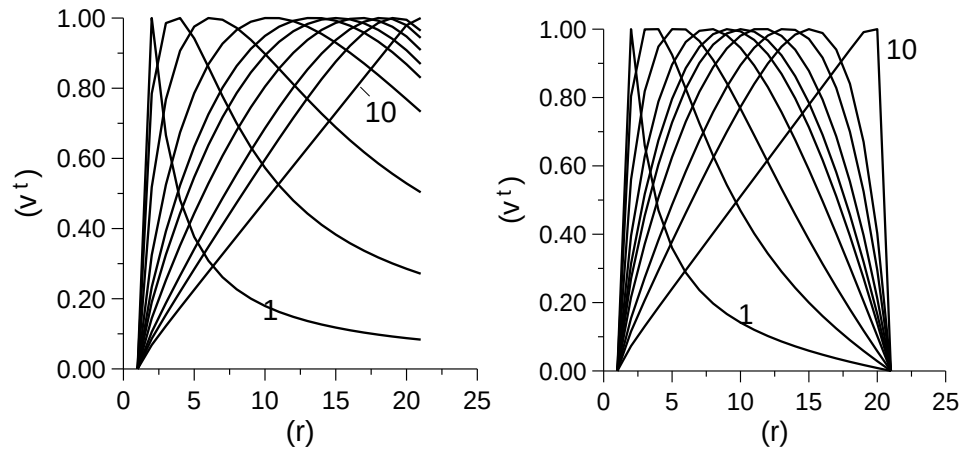

(a) (b)

Fig. 5. Initial conditions 1..10 concerning the transient part of the displacement for n=37 (a) Randolph-Wroth model, (b) Imre-Rózsa model.

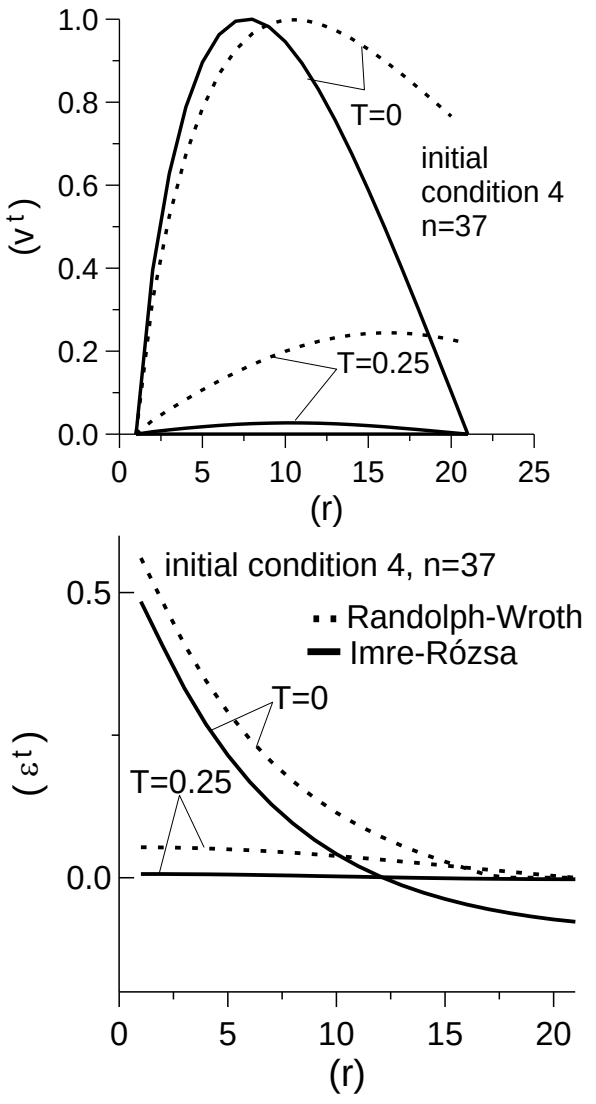

(a)

(b)

Fig. 6. (a) Variation of the displacement $v^t$ with t and r. (b) Variation of the volumetric strain $\varepsilon^t$ with t and r

The displacement at $r_1$ is constant for the Imre-Rózsa model (non-moving outer boundary). The results for the transient volumetric strain $\varepsilon^t$ show compression in the case of the Randolph–Wroth model. Partly compression (in the vicinity of the shaft) and, partly swelling (at the outer boundary) takes place in the case of the Imre-Rózsa model (Fig. 6). The pore water pressure dissipation is faster for the Imre-Rózsa model (Fig. 7a).

The result of the parametric analysis in terms of $K_0$ for some values for $K_0$ is shown in Fig. 7. According to the results, negative effective normal stresses were encountered within the displacement domain for both models for some $K_0$ values (Fig.7). It follows that hydraulic fracturing may occur, in accordance to the experiences [24, 26-27].

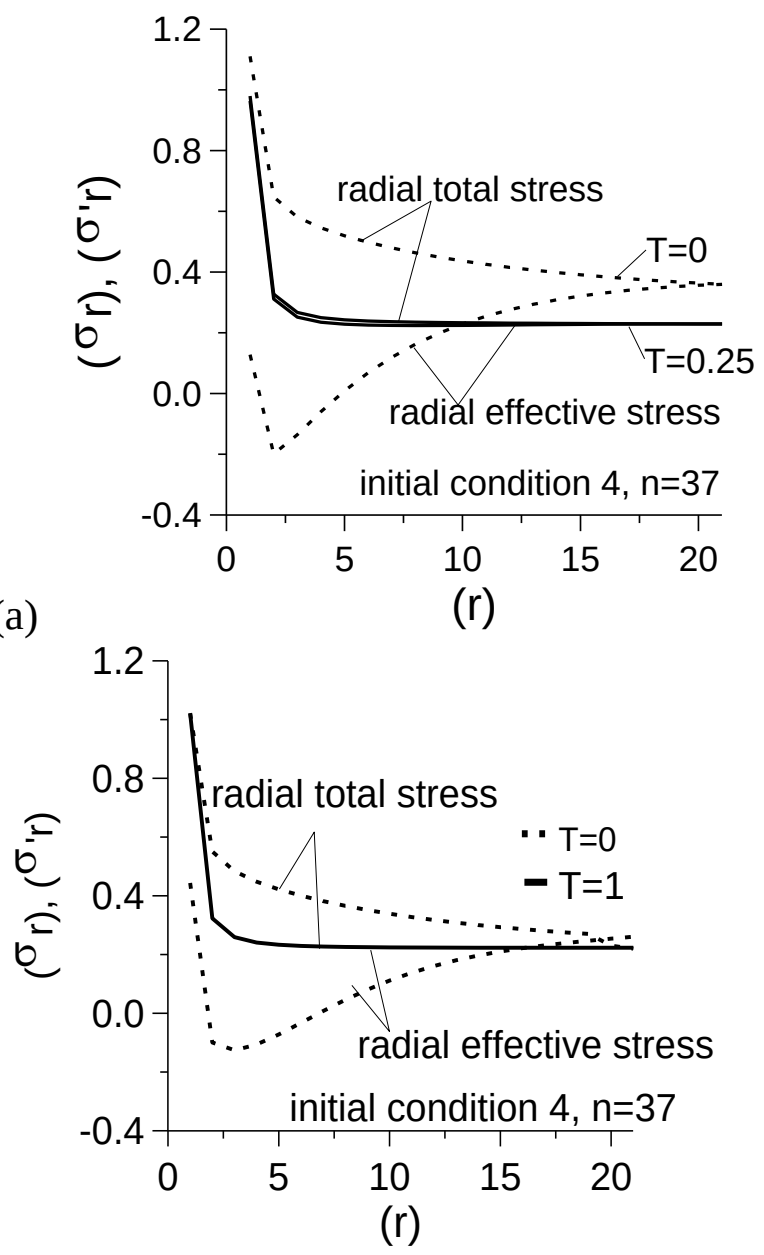

(a)

(b)

Fig. 7. (a) and (b). Time variation of the radial normal stresses. Imre-Rózsa model.and Randolph-Wroth model. Negative values are encountered in both cases.

# 6. Numerical properties

## 6.1. Methods

This chapter considers the numerical properties of the analytical solutions in terms of the space dimension, the initial condition and the boundary conditions.

It can be noted that the solution can be easier to be computed in the case m=1 or 3 using sin and cosine functions. The simplest numerical work is related to the oedometric models where no numerical solution is needed for the roots. The numerical analyses of the oedometric models is presented in [5]. For m=2, Bessel functions are needed which are approximated due to the slow convergence and, the so resulted series is 'semi-convergent', becomes divergent after a while.

The cylindrical cases with embedding space dimension $m$=2 were analysed for various displacement space domains and initial conditions. The space domain defined with radius $r_0$ =1.75 and $r_1$ = 64.75 are the case of undrained penetration, well above the tip, which was analysed in the previous chapter.

The boundary conditions were related to seven space domains with the same $r_0$ (Table 8). The precise roots of the boundary condition equations ($\lambda_i$, $\mu_i$) were determined with the secant method for the case of the $m$ =2 and 3 for the same of $r_0$ and seven values of $n$= $r_1$ / $r_0$, (Table 7). The number of terms was 40 at each specified $r_0$ and $r_1$ (some results for $m$ =2 and $n$=4 to 584 are shown see App A), 250 terms were considered for $n$=37. Ten values for the mean of the initial pore water pressure (parameter $D$) were selected, the value of parameter $F$ was determined for each value of $D$ and, for each value of $r_1$.

The initial pore water pressure functions are shown in Figure 2(a) in the case of $n$=37 and in Figure 2(b) for other $n$ values, where these were selected such that the D parameter was the same for the different $r_1$ values.

Table 8. Displacement domain geometry for the numerical tests (one value of $r_0$ and seven values of $n= r_1 / r_0$, $r_0$ =1.75 cm)

| Serial number of domains | $r_1$ [cm] | $r_1$- $r_0$ [cm] | $n=r_1/r_0$ |
|---|---|---|---|
| 1 | 7 | 5.25 | 4 |
| 2 | 33.25 | 31.5 | 19 |
| 3 | 64.75 | 63 | 37 |
| 4 | 127.75 | 126 | 73 |
| 5 | 255.5 | 253.75 | 146 |
| 6 | 511 | 509.25 | 292 |
| 7 | 1022 | 1020.25 | 584 |

## 6.2. Results

The aim is to determine the numerical properties of the analytical solution, which is related to two problems, is the computed series convergent at all and how many terms are needed to be used.

### 6.2.1. Computing Bessel function values

The initial condition is an infinite series of Bessel function, of the first and second kinds, order of 1 and 0, which converge very slowly for large values of the independent variable. According the usual practice ([25, 30]), these functions were approximated differently in the small (x<8) and large (x>8) range of the independent variable. The small range functions look like simple power laws and were approximated by rational functions. The large range functions look like sine or cosine with decay of $x^{-1/2}$. The products of polynomials and sine-cosine functions were used in the form of a library routine ([30]). The series applied in the range x>8 is not convergent in the sense of convergence of power series, after a certain number the terms begin to increase, even in the case of arbitrarily large x (semi-convergent series [25, 30]).

The separation of the two ranges can be seen in Figure 8 for the various space domains, noting that arguments are about the same for the two models on a given space domain. Concerning $r_0\lambda_k$ and $r_0\gamma_k$, large is the range if $k$>8 for the smallest space domain ($n$=4) and if $k$>90 for the space domain of the dissipation test ($n$=37). Concerning $r_1\lambda_k$ and $r_1\gamma_k$ they are in the large range if $k$>3 for the space domain of the dissipation test ($n$=37), and every k fall in the large range for $n$=4.

It follows that for $n$=37 the shaft stresses can be computed by the small range approximation precisely, at $r_1$ $k$~3 terms can be used in the small range approximation. In the case for small displacement domains (e.g. $n$=4) at $r_0$ $k$~8 terms can be used in the small range approximation, at $r_1$ the large range approximation has to be used which may entail some convergence problems. The few terms in the small range approximation may entail non-precise solution, the large range approximation may entail non-convergent series (see App B). Concerning the Imre-Rózsa model, the series after summation was convergent with $k$ up to about 200 terms then it became divergent for initial conditions 1 and 10. Concerning the R-W model, after around 125 terms the series became divergent for most initial conditions (Fig. 8). The solution is non-precise at the outer boundary for n=37 and the situation is worse for n<37.

### 6.2.2. Convergence and initial condition

The rate of convergence of the Fourier-Bessel expansion of the pore water pressure at $r=r_0$ was tested in the function of the initial condition shape functions for each seven space domain and model. The value of 1 was approximated as follows:

$$1=u_0(r_1,D,r_0)=\sum_{i=0}^{\infty} C_i\left\{\begin{array}{l}[I_0(\lambda_i r_0)+\mu_k Y_0(\lambda_i r_0)]\\ -[I_0(\lambda_i r_1)+\mu_k Y_0(\lambda_i r_1)]\end{array}\right\} \quad (66)$$

$$1=u_0(r_1,D,r_0)=\sum_{i=0}^{\infty} E_i[I_0(\gamma_i r_0)+\beta_i Y_0(\gamma_i r_0)] \quad (67)$$

Some fixed certain numbers the terms $k$ were considered. The results are summarized in the function of the initial conditions, space domain and number of terms in Figure 9, for the case of k<41, n=37. The error related to a certain $k$ is rapidly increasing as the mean initial condition ordinate $D$ varies 'towards' the $D$ value of the closest zero solution (i.e. $D\rightarrow 0$ for both models, $D\rightarrow 1$ for the Imre - Rózsa model). For the not too extreme initial conditions (i.e. 3 to 7), the numerical error is not important. This results can be explained as follows. If the initial condition series "converges" to the zero solution then the coefficients converge to zero, too. As a results, being any other term constant in Equations (52)-(53), the sum will decrease at every fixed $k$ for an initial condition 'being closer' to the zero solution. The decrease of the coefficients in terms of the initial conditions for any fixed k can be seen in Figure 10.

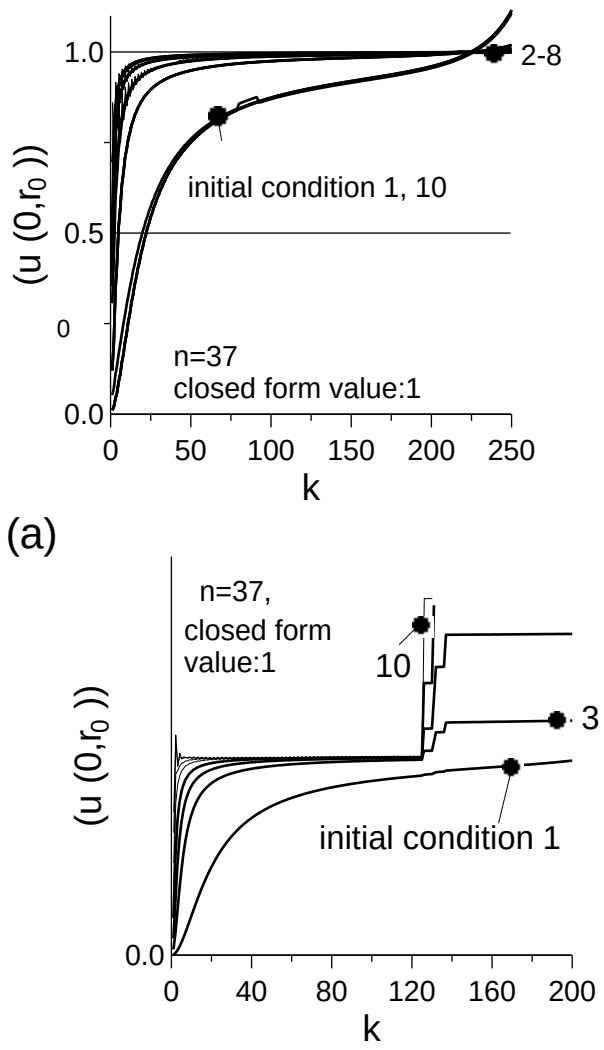

Fig. 8.The convergence, the influence of the initial condition and cut off number k. The $D$=0.001,0.016, ...0.970, according to Table 4. (a) coupled 1. (b) coupled 2. Note: The zero solution is closest to related to the constant initial conditions with $D$ =0 and 1 for the coupled 1 model, with $D$ =0 for the coupled 2 model.

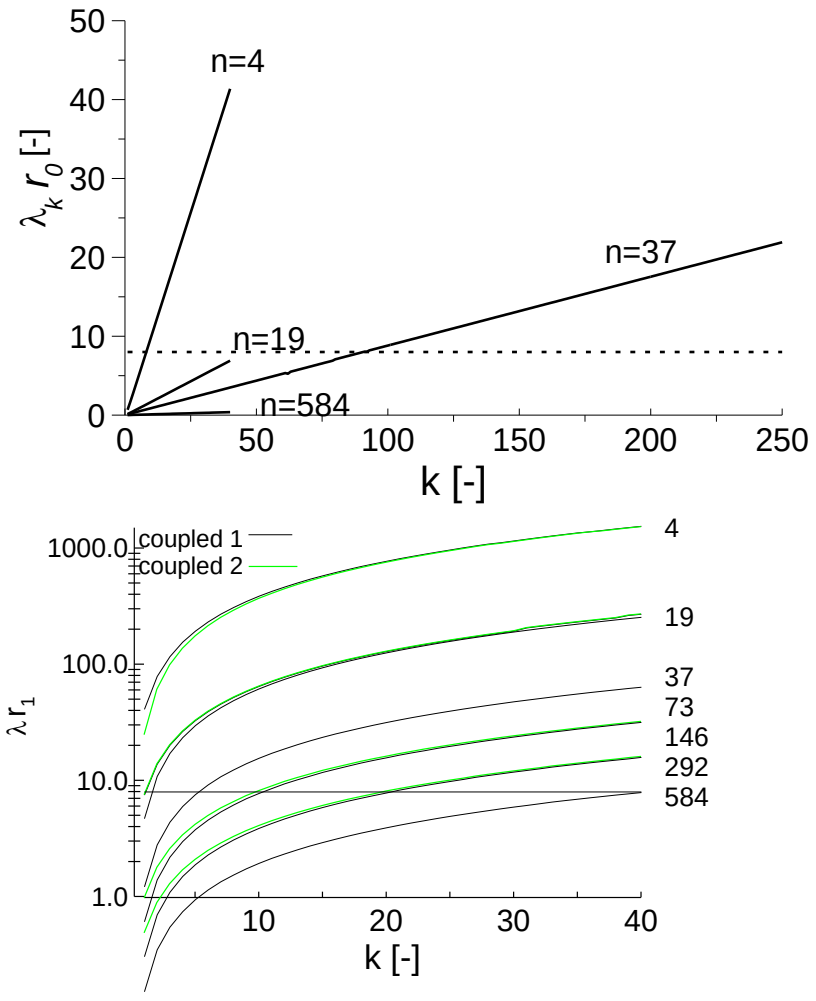

(a)

(b)

Fig. 9. (a) Bessel function approximation, cylindrical model (the analytical solution is acceptable below around 8. (b) Variation of the argument $r_0\lambda_k$ and $r_1\lambda_k$ with term number $k$ and displacement domain geometry number $n$. The solution is non-precise at the outer boundary for n=37.

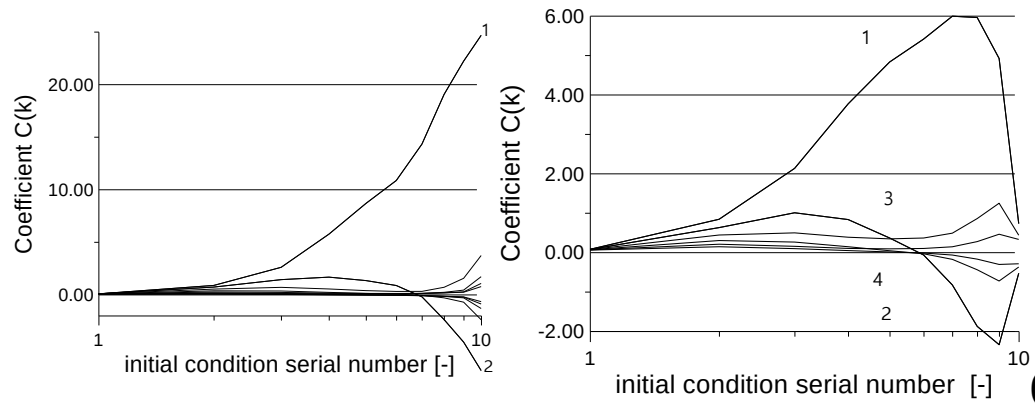

(b)

Fig. 10. The coefficients for fixed k, depending on the serial number Fig. 9. Cylindrical models. The Bessel series approximation of the initial pore water pressure on the shaft (i.e. with value of 1) after summation.

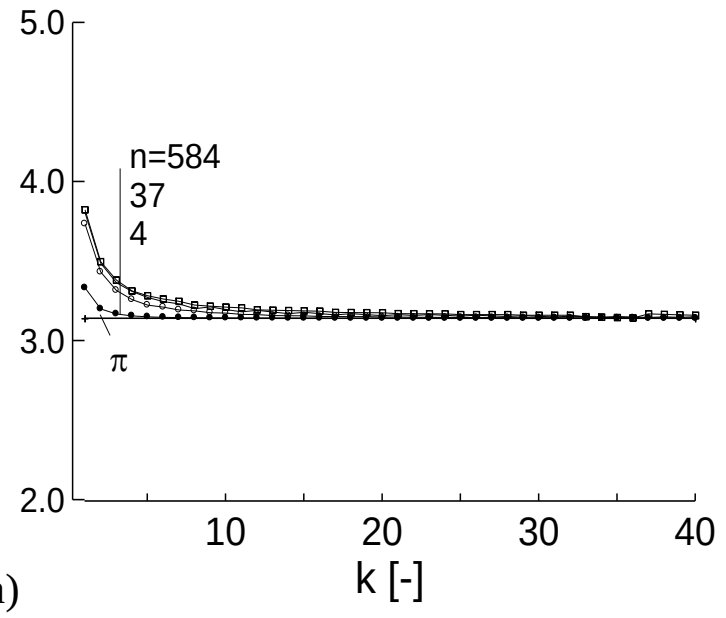

(a)

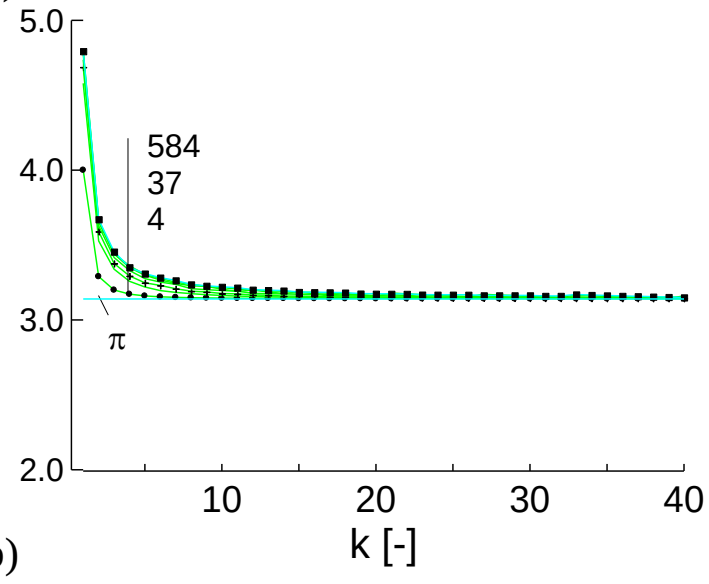

(b) (b)

Fig. 11. The validity of the approximate root formulae. . (a) (b) Results of the numerical tests for $\Pi=(r_1-r_0)\,\lambda_k/k$ or $\Pi=2(r_1-r_0)\,\gamma_k/(2k-1)$ using various values for $r_1$=n $r_0$ and $r_0$=1.75 cm.

## 6.2.3. Root formulae

In the case of the two cylindrical models ($m$ =2), 250 terms were considered for $n$=37 and 40 for other space domains. The precise roots of the boundary condition equations were determined with the secant method and, were transformed using the approximate closed form formulae. The results are shown in Figure 11. According to the results, the error of the approximate closed form formulae decreases with $k$ for both models. The validity of the approximate model law was numerically proven for one-term solution and for “non-extreme” initial conditions ([33]).

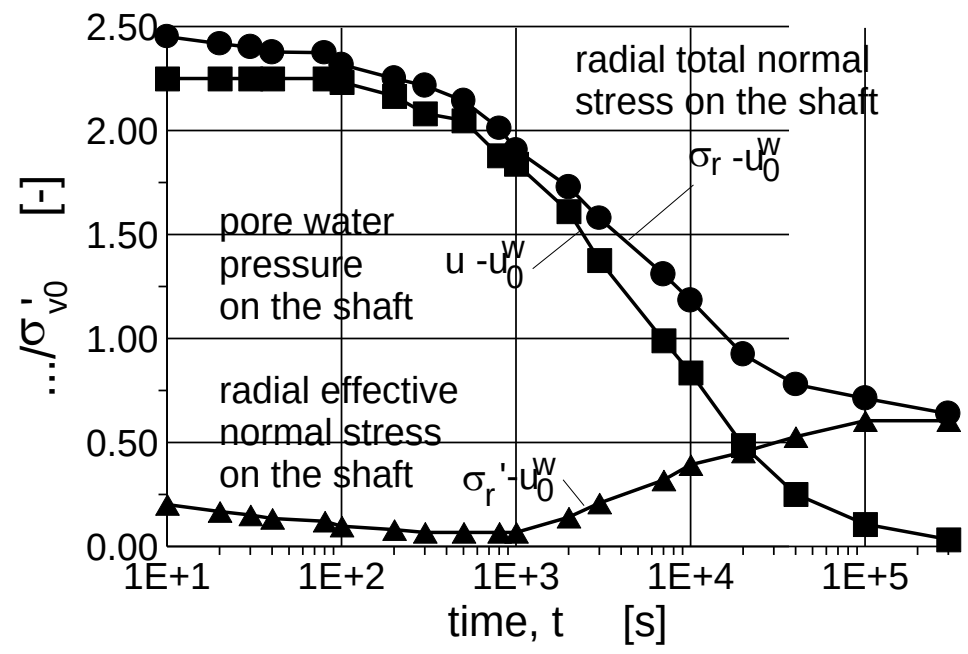

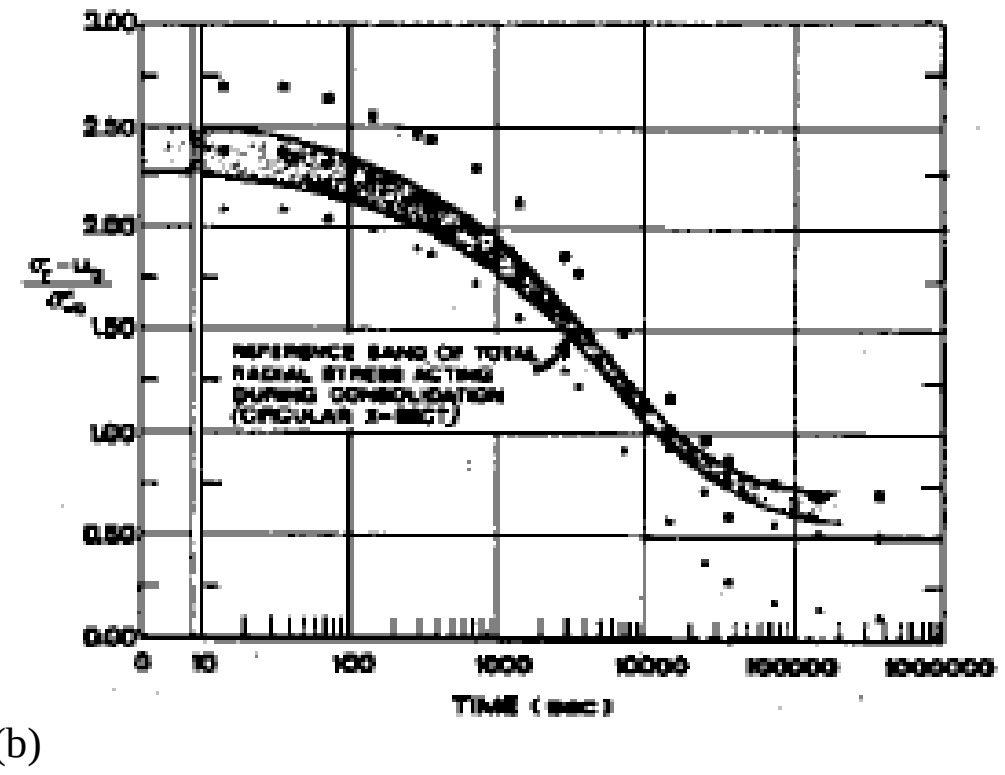

(b)

Fig. 12. Total stress dissipation test data in Boston Blue Clay. (a) The piezo-lateral stress cell test data. (b) DMTA data.

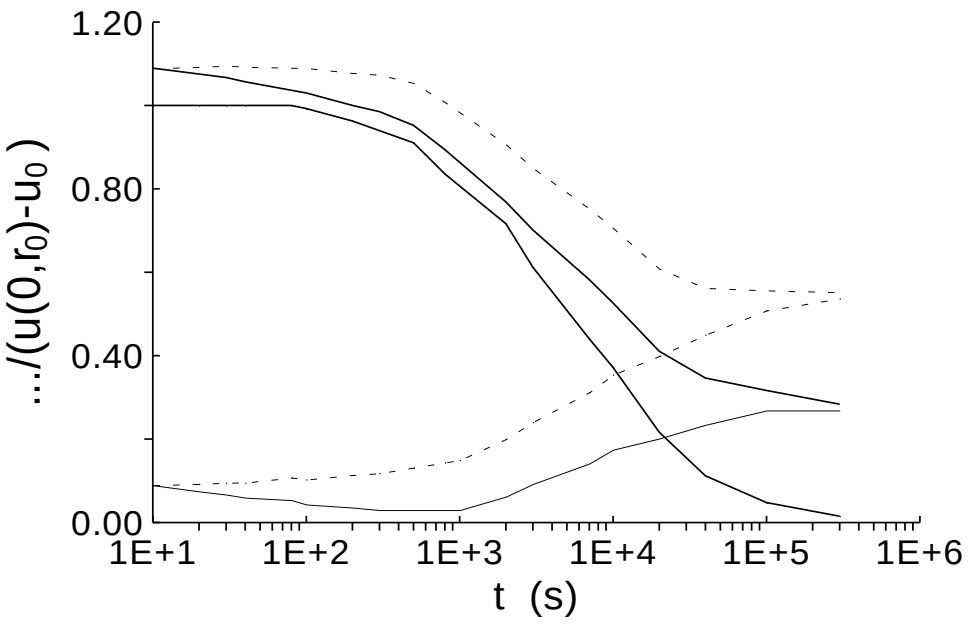

Fig. 13. Qualitative features of the suggested model family of cylindrical and oedometer models. Time dependent and time independent constitutive models, solid and dashed lines, resp., (precise and approximate, joined model and the linear consolidation model, the difference is the effect of the relaxation part-model).

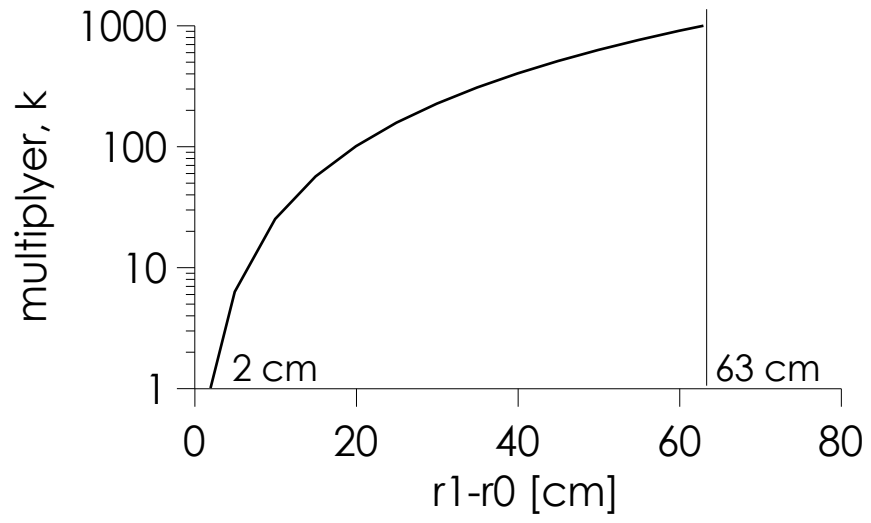

(c)

Fig. 14. (a) and (b) The displacement domain bounded by a 0-dimensional sphere (oedometer model), 1- dimensional sphere (cylindrical model). (c) The constant $k$ of the model law for the coefficient of consolidation $c_{cyl} = k\, c_{oed}$ (when CPT PSL test or DMTA data are evaluated with oedometer model where $r_1$-$r_0$=2cm, and cylindrical model where $r_1$-$r_0$=63cm.

## 7. Total stress dissipation tests

No mathematically precise method is used at present for interpreting the DMTA/CPT total stress dissipation curve (Fig. 12). Concerning the total and effective stress dissipation tests, the coupled 1 model - being completed by a relaxation term - was verified by measured oedometric relaxation test data (for unmoved boundaries [28-30]).
The total stress solutions of the coupled 1 models with various embedding space dimensions $m$=1 to 3 are qualitatively similar (Fig. 13). However, the controlling parameter (the mean initial normalized pore water pressure $D$) is significantly different for various the embedding space dimensions (i.e. for the linear initial function: $D \sim 1/(m+1)$).

In this work, the cylindrical coupled consolidation models with constant displacement boundary conditions were started to be used to evaluate some well-documented DMT data ($\sigma$). In this work the results of the evaluation with the oedometric models are also presented. The model law $c_{cyl} = k\, c_{oed}$ was derived from the time factor (Fig. 14).

The models were linear in the simplest form, the non-linear behavior was approximated by applying a relaxation part-model. Results are shown in Table 9, Figs 15 to 17.

According to the results, the identified c is slightly larger than expected. In other words, the CPTu total stress dissipation may entail 'too large' stress drop from modelling aspect possibly since the boundary is not unmoved due to the stress release (during the test, unclamped condition prevails) and, as a result, the diameter of the penetrometer may slightly decrease with time due to the stress release.

Table 9. Fucino DMT ($\sigma$) data at 5m, 10m, 15m. Cylindrical model, identified $c$ [cm$^2$/s] with confidence interval.

| | $c$ | $c_{max}$ | $c_{min}$ |
|---|---|---|---|
| 5m | 8,0E-3 | 1,0E-2 | 6,0E-4 |
| 10m | 8,0E-3 | 1,0E-3 | 6,0E-4 |
| 15m | 9,0E-3 | 2,0E-3 | 6,0E-4 |

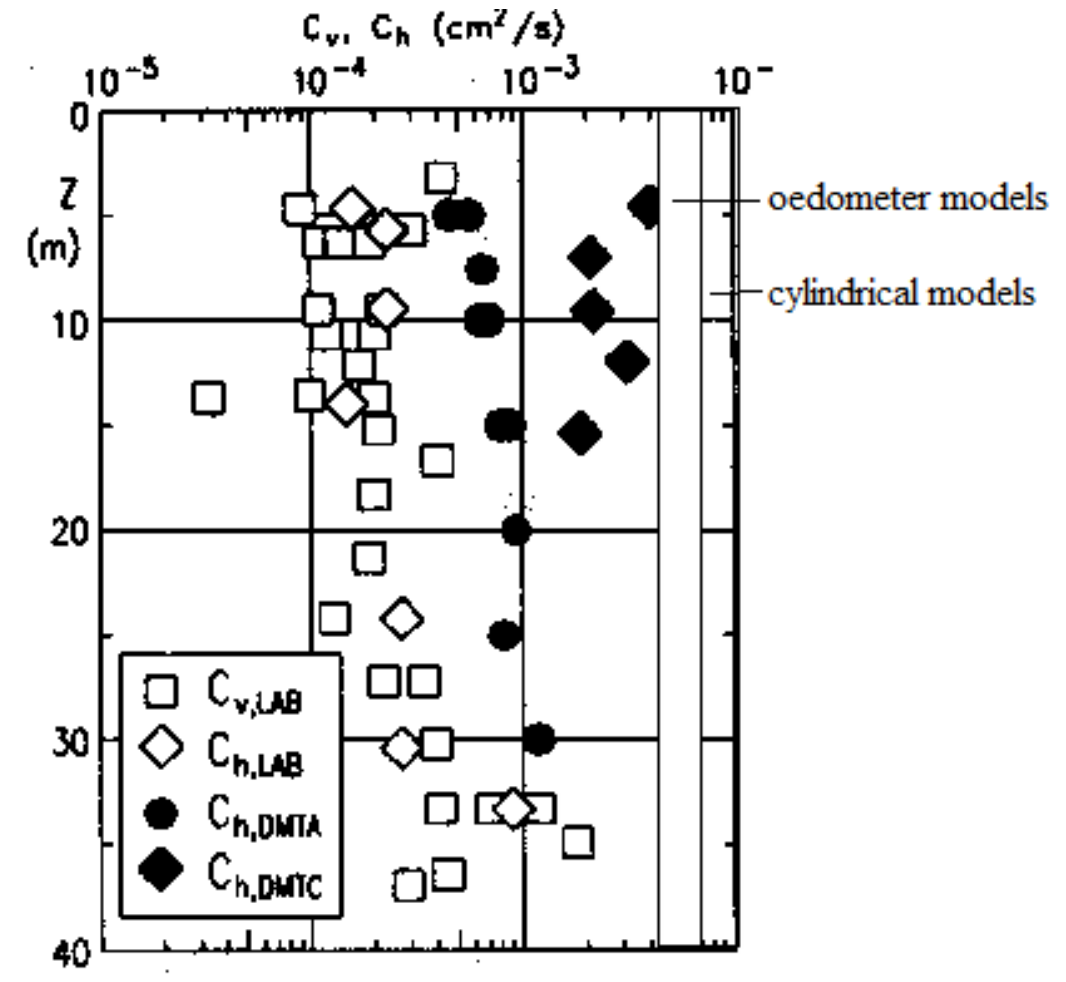

Fig. 15. Some identified $c$ values from this work, Fucino site. The model underpredicts the c values possibly due to the stress release effect.

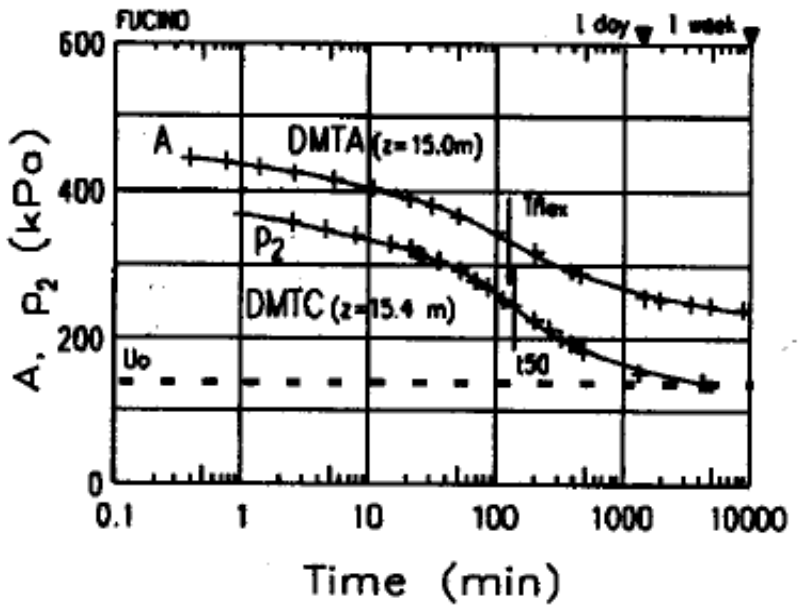

Fig. 16. Fucino DMTA and DMTC data measured at 5m. 10m, 15m (from [17]).

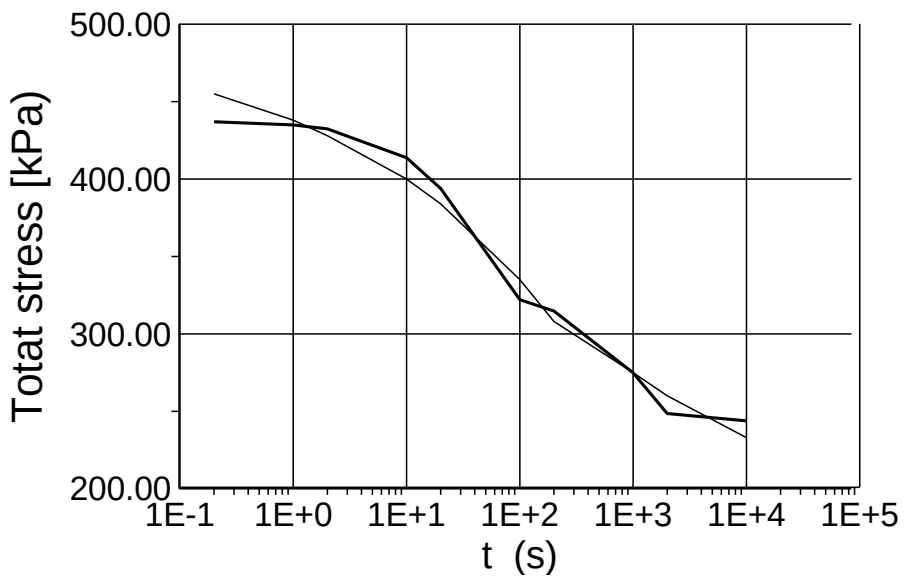

Fig. 17. Fucino 15 m, DMTA total stress data evaluation with cylindrical model 1a, 'fast method', fitted and measured data.

## 7.1. The pore water dissipation tests

At present a numerical solution of an uncoupled two-dimensional models is applied for approximate model fitting based on the use of the rigidity index which is estimated generally by thumb rules [15, 22, 31, 32].

In the frame of this research, both cylindrical models and the spherical coupled 1 model were included into an evaluation software for the evaluation of the CPTu tests, using a mathematically precise inverse problem solution method ([11] to [14]), needing the value of r1 (which may depend on the rigidity index) and its value is known for undrained penetration only. The point-symmetric 3 and the 2-dimensional coupled 1 solutions agree, the difference is tiny for typical initial conditions (Fig.18). The point-symmetric 2-dimensional coupled 1 and coupled 2 solutions agree at the physically admissible initial conditions (Figs. 2, 3c).

The analytical solutions are very useful in case of non-monotonic initial conditions, the dissipation curve is non-monotonic (Fig. 19) which may occur in CPTu testing if the soil is OC around the shaft (can originally be OC clay or can be a highly compressed sand or silt due to penetration which rebounds around the shaft). The modelling of non-monotonic dissipation curve is easier with the analytical models.

It can be noted that in case of partly negative initial pore water pressure distribution, the mean effective stress may initially increase during dissipation resulting in an initial total stress increase.

The analytical and numerical pore water pressure dissipation solution agree, as shown in Fig. 20 (related to Ir=150 and the conventional time factors).

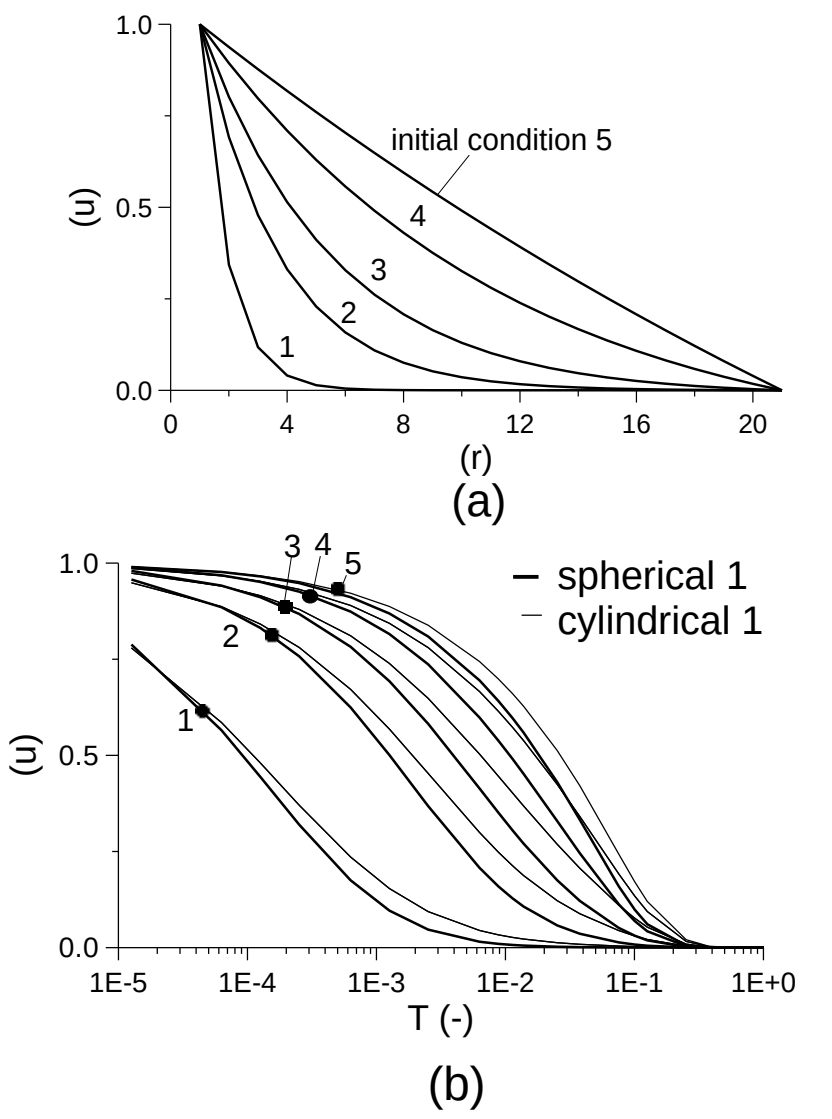

Fig. 18. Comparing coupled 1 dissipation curves in the function of the initial condition, spherical model solutiuons are faster.

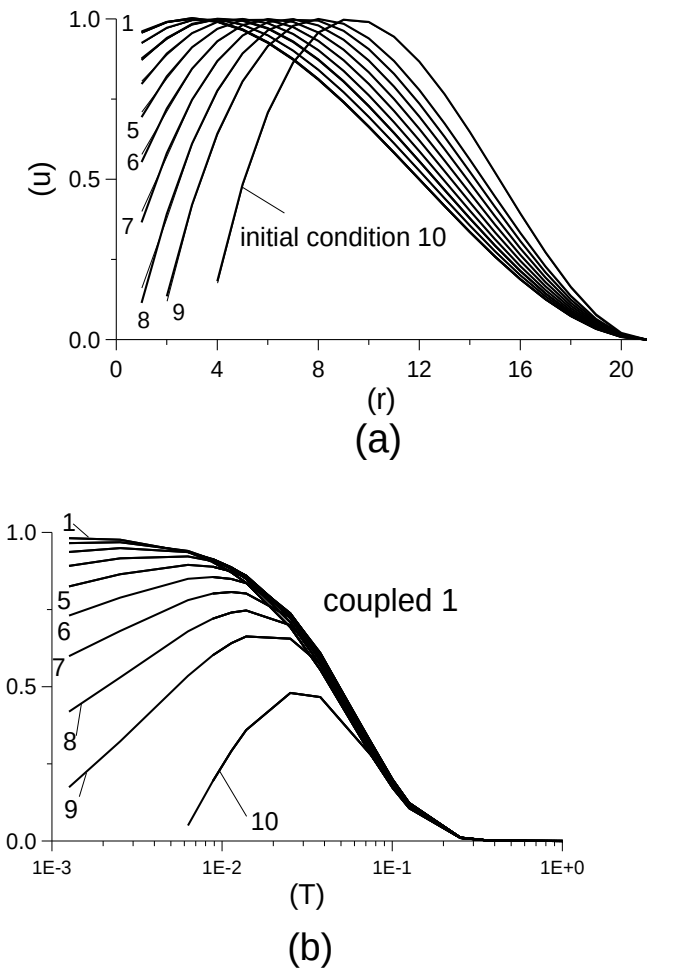

Fig. 19. Spherical dissipation curves, non-monotonic initial conditions result in non-monotonic disipation curves.

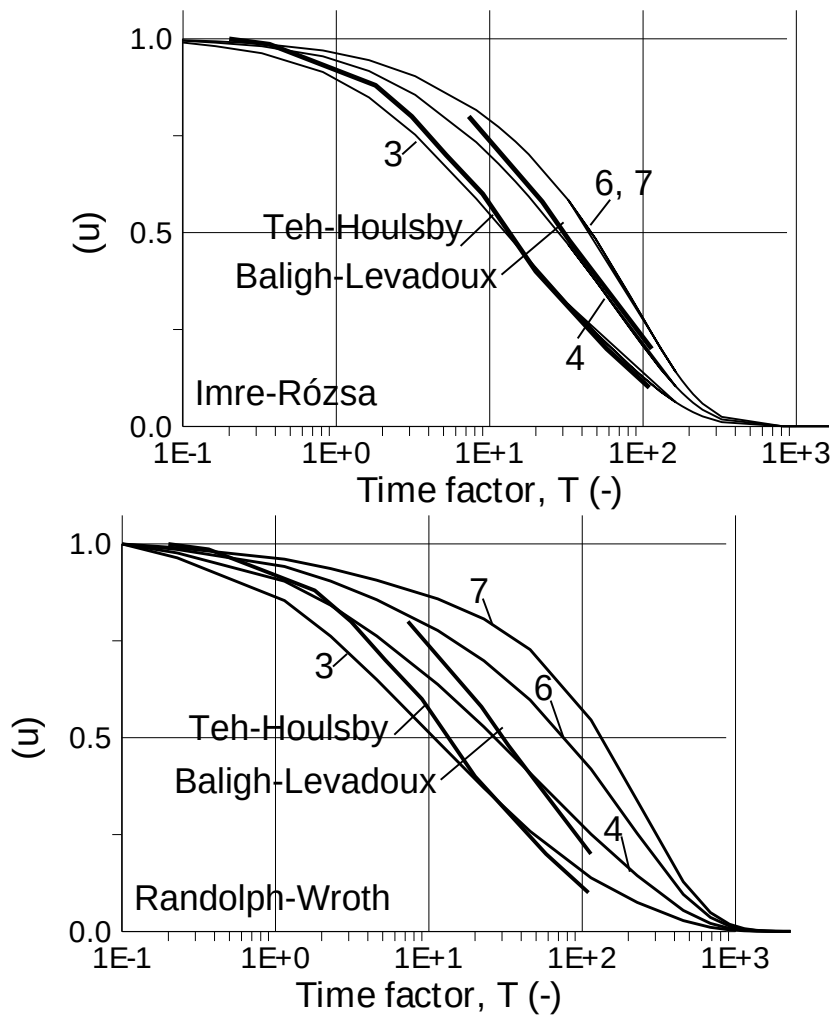

Fig. 20. Comparing the Imre-Rózsa model with two dimensional solutions (initial conditions 3 to 6). (b) Comparing the Randolph-Wroth model with two dimensional solutions (initial conditions 3 to 7).

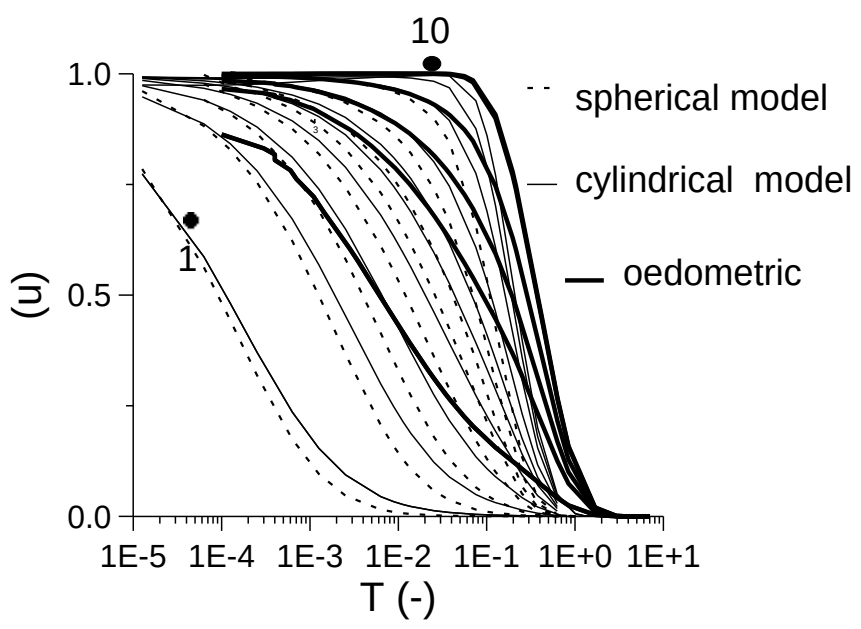

Fig. 21. Similarity of the dissipation curves, coupled 2 models. Spherical, cylindrical and oedometric models. Note: the tiny difference is related to the geometry of the displacement domain. The spherical model solutiuons are fastest, oedometric ones the longest.

# 8. Discussion

## 8.1. The analytical solution

### 8.1.1. Basic features

(1) The solution of each model from both coupled consolidation model family is equal to the sum of the solution of the steady-state seepage model (being zero here), the solution of the drained continuum-mechanical model, the solution of the transient seepage model and the self-weight stresses.

(2) The drained continuum-mechanical model is the oedometer ($K_0$) compression model for m=1 and the cavity expansion models for m=2, 3, with two kinds of boundary conditions.

(3) The transient solution part has the form of Bessel series with order of m/2 and (*m*-2)/2. The coefficients can be determined from the initial condition, the constants can be determined from the boundary conditions.

(4) The transient effectives stresses are negative in the vicinity of the shaft, the sum of the steady-state and transient effective stresses may become negative, implying the possibility of hydraulic fracturing. This may occur if the steady-state normal stress may become small due to a rebound, after partial unloading.

(5) The dissipation is instantaneous at the shape function of related to D =1 (constant, non- zero function) for the coupled 1 model since in the case initial, transient effective stress state is identically equal to zero. It follows that the intensive variable of the coupled seepage in soils is the effective stress, the stress state variable the saturated soils.

### 8.1.2. Computing the analytical solution

(1) The constants from the boundary condition equations can be computed with a closed-form formula only for embedding space dimension m=1 and cannot be computed with closed form formulae for m=2 and 3. In the latter case the constants were determined by the secant method numerically for various displacement domains.

(2) For m=2 and 3, by using the asymptotic Bessel formulae, two approximate, closed-form root formulae

were derived, being the same as the precise formula of the one dimensional case within a model family.

(3) As a result, the transformation of the approximate solution on the displacement domain is possible (model law). The suggested time factors are differing by a constant multiplier from the one usually applied:

$$T = \frac{ct}{(r_1 - r_0)^2} = \frac{ct}{r_0^2\left(\frac{r_1}{r_0}-1\right)^2} \quad \text{or}$$

$$T = \frac{ct}{4(r_1 - r_0)^2} = \frac{ct}{4r_0^2\left(\frac{r_1}{r_0}-1\right)^2} \quad (68)$$

(4) Having identical (approximate) dimensionless time coordinate, it was possible to compare the pore water pressure dissipation solutions for all embedding space dimensions within a model family (Fig. 19). Due to the similarity, the solutions of the m=1, 2 and 3 solutions can be interchanged, the solutions of the m=1or 3 model can be used instead of the m=2 models in the evaluation. The differences in the identified parameters for the various models can likely be compensated by constant multiplyers. (The dissipation is the faster if the domain has larger boundary, is more 'rounded'.)

(5) The steady-state part of the solution was determined experimentally, the sum of the gravitational stresses and the cavity expansion stresses were taken from real-life PLS data. The analytical solution of the coupled 1 and 2 models clearly indicated locally negative values for the effective stresses inside the displacement domain in agreement with the experiences (hydraulic fracturing around sea-bad wells, around piles [30] and in the oedometric relaxation test in case of load reversal [23]).

### 8.1.3. Convergence

The convergence of the analytical solution depends on the initial condition in the same way within each model family independently of the *m* embedding space dimension values (oedometer:1, spherical:3, cylindrical:2). Since the coefficients vary continuously with initial condition parameter as $D\to 0$ to zero, due to the small-valued coefficients, larger number of terms may give similar accuracy close to a zero solution. Far from the zero solutions, for the not extreme initial conditions (i.e. 3 to 7, Fig. 2), the 10 to 40 -term approximation gives good accuracy, however, the error rapidly increased getting 'closer' to a zero solution (i.e. $D\to 0$ for both models, $D\to 1$ for the coupled 1 model).

Due to the slow convergence, a Bessel series approximation is needed in case of the cylindrical model ($m$=2). Assuming undrained penetration ($n= r_1\, r_0$ =37), the shaft stresses can be computed with acceptable preciseness using the small range approximation. At $r_1$ the large range approximation has to be used which may entail some convergence problems. Concerning the Imre-Rózsa model, the series after summation is convergent with $k$ up to about 200 terms, being le then it became divergent for initial conditions 1 and 10.

Assuming that the size of the displacement domain decreases ($n=r_1/\ r_0$ <37) in case of partly undrained penetration with increasing soil permeability to $r_1\to r_0$, then preciseness of the shaft stress approximation decreases with decreasing $r_1$ and decreasing $D$ and becomes not acceptable at around $n=r_1/\,r_0$ =4.

Computing the value at $r=r_0$ of the normalized initial condition shape functions, the numerical convergence test resulted in the same picture within a model family. For the not extreme initial conditions (i.e. 3 to 7, Fig. 2), the 10 to 40 -term approximation gave good accuracy, however, the error rapidly increased getting 'closer' to a zero solution (i.e. $D\to 0$ for both models, $D\to 1$ for the coupled 1 model) since the coefficients of the Bessel series decreased rapidly.

(6) In the cases m=1 and m=3 the solution can be computed by using sin and cosine functions. However, for m=2, the Bessel function series had to be computed. Due to the low convergence, the series were approximated, according to the practice ([26]), differently in the small (x<8) and large (x>8) range of the independent variable. The preciseness in the small range was acceptable, in the large range the series was not convergent in the sense of convergence of power series.

It was found that in case of n=37 the series was convergent in the small range up to about 90 terms at $r=r_0$. The n=37 series was in the small range only up to about 8 terms at $r=r_1$ and the n=4 series was in the small range up to about 7 terms at r= r0, the result was realistic but non-precise with 40 terms in these cases.

## 8.2. Model family features

### 8.2.1. Analysis of the model

The solution of the system of Equations (1) and (2), the initial and the boundary conditions were analysed assuming an initial pore water pressure function series $u_0(r)$ of positive and monotonic function, being in unique relation with a single parameter (i.e. the initial mean pore water pressure, denoted by $D$). The main features of the transient part of the solution of the two model-families are as follows.

1) The transient effective stress depends differently on the pore water pressure function (and, on the initial pore water pressure distribution) for the coupled 1 and for the coupled 2 models, respectively, being equal to **($u_{mean}$-$u$) and –$u$**, respectively. For the coupled 1 models the mean effective stress and the volume of the domain is constant during consolidation, swelling takes place in the vicinity of the shaft, compression takes place in the vicinity of r1. For the coupled 2 models, the domain is decreasing by moving the boundary r1 inward during consolidation. The mean effective stress is increasing by the value of the initial mean pore water pressure.

2) The transient radial total stress depends differently on the pore water pressure function (and, on the initial

pore water pressure distribution) for the coupled 1 and for the coupled 2 models, respectively, being equal to $\boldsymbol{u_{mean}}$ **and 0**, respectively (i.e. the mean pore water pressure and the zero function). As a result, for positive pore water pressures, the total stress is decreasing at the shaft – soil interface by the value of the initial mean pore water pressure for the coupled 1 model during dissipation. For the coupled 2 models, at the shaft – soil interface and the total stress is constant.

It can be noted that in case of partly negative initial pore water pressure distribution, the mean effective stress may initially increase during dissipation resulting in an initial total stress increase.

3) The integral of the dissipation curve on the time domain is a functional depending on the initial, transient effective stress state, being different for the coupled 1 than the coupled 2 models. This functional characterizes the rate of consolidation (or seepage).

As a first consequence, the effective stress (and the functional) is smaller (the pore water pressure dissipation is faster) for the coupled 1 than for the coupled 2 model for any fixed initial condition.

The effective stress is the zero function (the functional is zero) at the zero initial pore water pressure function for both model-families (this is the trivial zero solution, at $D$=0) and at the constant, non-zero initial pore water pressure function for the coupled 1 models (this is a non-trivial zero solution, at $D$=1).

As a second consequence, at the non-trivial zero solution, the dissipation is instantaneous. As a result, if the initial condition of the oedometer relaxation test contains a constant component, this will dissipate instantaneously. As a third consequence, if the initial condition is the closer to one of a zero solution (eg., in terms of initial condition parameter $D$), then the dissipation is the faster.

## 8.2.2. Similarity of solutions

Having identical (approximate) dimensionless time coordinate, it was possible to compare the pore water pressure dissipation solutions for all embedding space dimensions within a model family (Figs. 18, 21).

According to the results, the dissipation is the slightly faster if the domain has larger boundary, is more 'rounded', but the similarity is surprising.

Due to the similarity, the solutions of the m=1, 2 and 3 solutions can likely be interchanged, the solutions of the m=1or 3 model can be used instead of the m=2 models in the evaluation.

The slight differences in the identified parameters for the various models can likely be compensated by some constant multiplyers. Further research is suggested on the similarity.

Concerning the two model families, the dissipation curve solutions nearly agree, the (newly introduced) time factor values are increasing slightly with decreasing embedding space dimension due to geometrical reasons (Fig. 15).

## 8.3. Physics features

## 8.3.1. Note on the uncoupled models

It can be examined whether or not the uncoupled model can be derived from the coupled 2 (i.e. Randolph-Wroth model). The integral expressions and, the constitutive law are used to express the radial total stress and, the first invariant of the total stress tensor:

$$\sigma_r = \frac{\mu}{1-\mu}\sigma_0 + \frac{1-2\mu}{1-\mu}u \quad (69)$$

$$\sigma_I = \frac{1+\mu}{1-\mu}\sigma_0 + 2\frac{1-2\mu}{1-\mu}u \quad (70)$$

in terms of the pore water pressure. This may be constant if the μ - the Poisson's ratio in terms of the effective stress – is equal to 0.5. This case is impossible (soil is incompressible and there is no consolidation). It follows that the uncoupled model can not be derived from the coupled model.

## 8.3.2. Thermodynamic interpretation

The transport of any extensive quantity implies an intensive quantity the homogeneous distribution of which is the precondition of the equilibrium (Theorem 0 of Thermodynamics). The movement of an extensive quantity is caused by the inhomogeneous distribution of the intensive quantity which is tended to be eliminated (Law 2 of Thermodynamics).

The extensive variable for seepage is the water mass or volume. The intensive variable for seepage is the total hydraulic head of the water phase:

$$h = z + \frac{u}{\gamma_w} \quad (71)$$

where z is the vertical distance from an arbitrary datum. In the models presented here the effect of $z$ was neglected assuming that $h=u/\gamma_W$.

The rate of the dissipation at any point $r$ – being characterised by the area of the subgraph of the dissipation curve $u(t,r)$ –was expressed as a functional of the initial transient volumetric strain. It follows from this expression that the dissipation is instantaneous if the initial transient volumetric strain is identically equal to zero. It may occur that the initial transient volumetric strain is identically equal to zero while the initial pore water pressure function is non-zero (i.e. coupled 1 model, constant initial pore water pressure distribution). The dissipation of the pore water pressure is instantaneous in this case. Therefore, it can be said that transient seepage takes place if and only if the initial transient volumetric strain is not identically equal to zero.

## 9. Conclusions

Two linear, point-symmetrics coupled consolidation model families, with embedding space dimensions $m$=1 to 3, differing in the boundary condition at the outer, boundary.

Constant displacement boundary condition is assumed at the inner boundary, which is at $r=r_0=0$ for the two kinds of oedometer tests, being the symmetry point of a double-drained oedometric samples. The outer boundary $r=H$, where zero pore water pressure boundary condition is assumed besides the total stress load (coupled 2 models) or displacement load (coupled 1 models).
The $r=r_0$ boundary is the surface of the model pile. The outer boundary $r=r_1$, where zero pore water pressure boundary condition is assumed, is unknown. It is determined as the zero pore water pressure line after penetration. The coupled 1 models assume constant displacement, the coupled 2 models assume constant volumetric strain here. The main points of the results are repeated here as follows.

1) The differences between the two model families are significant as follows, in the total stress modelling. The coupled 1 models describes the total stress drop during disspation (and the effective stress variation) encountered during the CPT or DMT total stress dissipation tests, the coupled 2 models prognosticate constant total stress at the shaft-soil interface.

2) The similarity of the solution within a model family is surprising, and the modeified Terzaghi‘s model law can be used to model eg. DMT dissipation test by an oedometer test evaluation model. The analytical and numerical properties are similarly dependent on the initial condition and the numerical properties are worse closer to the zero solutions in each case.

3) Being similar, the analytical solutions can be interchanged within a model family (ie., in other words, coupled 1 models for m= 1, m= 2 and m=3 can be interchanged). In addition, the oedometer tests can be used to study the phenomena after pile penetration (eg., to study the effect of the stress release of the elastic pile material after pile penetration).

4) The coupled 1 modelling was verified for m=1 oedometer test case and started to be adalysed for the DMT dissipation test. First result indicate that in the latter case the stress release may influence the boundary condition.

5) The steady-state part of the solution may influence the effective stress. Using experimental data, the computed effective stresses can be negative locally inside the displacement domain in agreement with the experiences. (Hydraulic fracturing occurred several times around sea-bad wells or around piles after penetration ([30] and in the oedometric relaxation test in case of load reversal [28]).

6) The analytical solutions is numerically simple in the case m=1 or 3 being sin and cosine functions. For m=2, Bessel functions are needed with slow convergence and, which are approximated. The so resulted series is ‘semi-convergent’, being divergent for some space variables eg., in the case of small displacement domains.

7) Very few pieces of information are available on the relaxation (time dependency of constitutive law) which is needed to be considered for the total stresses around piles.

8) Very few pieces of information are available on the initial condition of partly drained penetration and it follows from the results of the analyses of the numerical properties of the analytical solutions that the dissipation modelling after partly drained penetration may face some numerical difficulties.

## 10. APPENDIX A

### 10.1. Derivation of the system of partial differential equations

The system of differential equations can be derived as follows for $n$=1..3 ($n$ = 1, 2, 3, [20]):

Equilibrium equation:

$$\frac{\partial \sigma_r}{\partial r} + \frac{(n-1)(\sigma_r - \sigma_\varphi)}{r} = 0 \quad (72)$$

Continuity equation:

$$\frac{\partial q}{\partial r} + \frac{\partial \varepsilon}{\partial t} = 0. \quad (73)$$

The generalized Darcy's law:

$$\frac{\partial q}{\partial r} = -\frac{k}{\gamma_v} \Delta\, u. \quad (74)$$

where the Laplacian operator:

$$\Delta = \frac{1}{r^{n-1}} \frac{\partial}{\partial r}\left( r^{n-1} \frac{\partial}{\partial r}\right) \quad (75)$$

The effective stress equality:

$$\sigma_{ii} = \sigma'_{ii} + u. \quad (76)$$

The geometrical equations:

$$\varepsilon_r = \frac{\partial v}{\partial r} \quad (77)$$

$$\varepsilon_{\varphi i} = \frac{v}{r} \quad (78)$$

$$\varepsilon = \frac{\partial v}{\partial r} + (n-1)\frac{v}{r} \quad (79)$$

or

$$\varepsilon = \frac{1}{r^{n-1}} \frac{\partial}{\partial r}( r^{n-1} v) \quad (80)$$

The constitutive equations:

$$\sigma'_r = \frac{-2G}{1-2\mu}\left[ (1-\mu)\frac{\partial v}{\partial r} + (n-1)\mu\frac{v}{r}\right] \quad (81)$$

$$\sigma'_{\varphi i} = \frac{-2G}{1-2\mu}\left[ \mu\frac{\partial v}{\partial r} + [(1-\mu)+(n-2)\mu]\frac{v}{r}\right] \quad (82)$$

The equilibrium equation, the effective stress equality, the geometrical and the constitutive equations are combined to give the modified Equilibrium Equation. The continuity equation, the Darcy's law and the geometrical equation are combined to give the modified Continuity Equation.

### 10.2. Solution

The general solution of the models, subject to the specified boundary conditions, is the sum of two parts: one transient and one steady-state. The steady-state part of the displacement ($v^p$) is given by the solution of the following equation (part of the modified Equilibrium Equation):

$$E_{oed} \frac{\partial \varepsilon}{\partial r} = 0 \quad (83)$$

This is the cavity expansion model for $n$=2 and 3, and the $K_o$ compression model for $n$=1. The solution has the following general form:

$$v^p = \frac{\alpha}{r^{n-1}} + \beta\, r \quad (84)$$

where the parameters $\alpha$ and $\beta$ can be determined from the non-homogeneous boundary conditions.

The steady-state pore water pressure solution is the solution of the Laplacian Equation (part of the modified Continuity Equation) which is equal to zero here.

The transient solution parts for the volumetric strain ($\varepsilon^t$), the displacement ($v^t$) and the pore water pressure ($u$), respectively(see App 2):

$$v^t(t,r) = r^{\frac{-(n-2)}{2}} \sum_{k=1}^{\infty} C_k \,[\, J_{n/2}(\lambda_k r) + \mu_k Y_{n/2}(\lambda_k r)]\, e^{-[\lambda_k]^2 ct} \quad (85)$$

$$\varepsilon^t(t,r) = r^{\frac{-(n-2)}{2}} \sum_{k=1}^{\infty} C_k \lambda_k \left\{ [\, J_{(n-2)/2}(\lambda_k r) + \mu_k Y_{(n-2)/2}(\lambda_k r)]\right\} e^{-[\lambda_k]^2 ct} \quad (86)$$

where $J_p$ and $Y_p$ are the Bessel functions of the first and second kind, of order $p$, $n$ is the space dimension; $\lambda_k$, $\mu_k$ are the roots of the boundary condition equations (composed from the homogeneous form of the boundary conditions); $C_k$ ($k$=1...∞) are the Bessel coefficients determinable from the initial condition, and $c$ is coefficient of consolidation ($c= k\ E_{oed} /\gamma_v$). Around 250 roots for constants $\lambda_k$, $\mu_k$ were determined for the models (see [8]).

The pore water pressure is determined from $v^t$ by integrating the equilibrium Modified Equilibrium Equation with respect to $r$ including boundary condition *Nr.* 1:

$$u^t = E_{oed} \int_{r_1}^{r} \frac{\partial \varepsilon^t}{\partial r} = E_{oed} \left[\varepsilon^t - \varepsilon^t(r_1)\right] \quad (87)$$

### 10.3. Analytical solution of transient displacement equation

A unified derivation of the analytical solution is presented here in the function of the space dimension.

The displacement equation:

$$-c\frac{\partial}{\partial r}\left[ r^{n-1} \frac{\partial}{\partial r}\left(\frac{1}{r^{n-1}}\frac{\partial}{\partial r}( r^{n-1} v )\right)\right] + \frac{\partial}{\partial t}\frac{\partial}{\partial r}( r^{n-1} v ) = 0 \quad (88)$$

The solution of the displacement equation can be expressed in the form of the product of $R(r)T(t)$, resulting in the equations:

$$-cT\frac{\partial}{\partial r}\left[ r^{n-1} \frac{\partial}{\partial r}\left(\frac{1}{r^{n-1}}\frac{\partial}{\partial r}( r^{n-1} R )\right)\right] + \dot{T}\frac{\partial}{\partial r}( r^{n-1} R ) = 0 \quad (89)$$

$$\frac{\dot{T}}{T} = \frac{c\frac{\partial}{\partial r}\left[ r^{n-1} \frac{\partial}{\partial r}\left(\frac{1}{r^{n-1}}\frac{\partial}{\partial r}( r^{n-1} R )\right)\right]}{\frac{\partial}{\partial r}( r^{n-1} R )} = -\alpha \quad (90)$$

$$\frac{c\frac{\partial}{\partial r}\left[ r^{n-1} \frac{\partial}{\partial r}\left(\frac{1}{r^{n-1}}\frac{\partial}{\partial r}( r^{n-1} R )\right)\right]}{\frac{\partial}{\partial r}( r^{n-1} R )} = -\alpha \quad (91)$$

$$\frac{T'}{T} = -\alpha \quad (92)$$

The R equation has a solution

$$R_1 = \frac{1}{r^{n-1}} \quad (93)$$

If another solution is expressed in the form

$$R_{2,3} = \frac{1}{r^{n-1}} w(r) \quad (94)$$

then the following can be derived assuming

$$u(r) = \frac{\partial w(r)}{\partial r} \quad (95)$$

$$u'' - u'\frac{n-1}{r} + \left(\frac{n-1}{r^2} + \frac{\alpha}{c}\right) u = 0 \quad (96)$$

which is the solution of the differential equation:

$$y'' + \frac{A}{x} y' + \left( B x^E + \frac{D}{x^2}\right) y = 0 \quad (97)$$

on the space domain [0,1] (see e.g. Kármán and Biot ([45]); Rózsa ([46])):

$$y = x^{\alpha\alpha}\ J_\nu(k x^\gamma) + x^{\alpha\alpha} Y_\nu(k x^\gamma) \quad (98)$$

where $J_v$, $Y_v$ are Bessel functions of the first and second kind, and of order $\nu$ and

$$A = 1 - 2\alpha\alpha,\ B = k^2\gamma^2,\ E = 2\gamma - 2,\ D = \alpha\alpha^2 - \nu^2\gamma^2 \quad (99)$$

Considering these, we find that

$$A = -(n-1),\ B = \alpha / c,\ E = 0,\ D = n-1 \quad (100)$$

$$\alpha\alpha = \frac{n}{2},\ k = \sqrt{\alpha / c},\ \gamma = 1,\ \nu^2 = [(n-2)/2]^2 \quad (101)$$

Therefore

$$u = r^{\frac{n}{2}} \left[ J_{\frac{n-2}{2}} ( \sqrt{\alpha / c}\ r ) + Y_{\frac{n-2}{2}} ( \sqrt{\alpha / c}\ r ) \right] \quad (102)$$

and the solution for the displacement equation can be computed from this by integration.

$$R_{2,3} = \frac{1}{r^{n-1}} \int u(t,r) dr = \frac{1}{r^{n-1}} r^{n/2} \left[J_{n/2} ( \sqrt{\alpha / c}\ r ) + Y_{n/2} ( \sqrt{\alpha / c}\ r )\right] \quad (103)$$

Changing the parameter notation, the displacement solution can be written as follows:

$$v^t(t,r) = \sum_{k=1}^{\infty} \left\{ A_k r^{-(n-1)} + r^{-\frac{(n-2)}{2}} C_k [ J_{n/2}(\lambda_k r) + \mu_k Y_{n/2}(\lambda_k r)] \right\} e^{-[\lambda_k]^2 ct} \quad (104)$$

where the constants $A_k$, $C_k$, $\lambda_k$ and $\mu_k$ can be determined from the boundary conditions and the initial condition.

### 10.4. Constants of the solution

#### 10.4.1. Initial condition

Using the initial condition for the transient displacement $v^t_o$ ($r$), the coefficients $C_k$ ($k$=1...∞) - the Bessel coefficients – can be determined using the orthogonality of the Bessel function series expansion.

From $v^t_o$ ($r$) the initial pore water pressure distribution $u^t_o$ ($r$) can be determined using Modified Equilibrium Equation and vice versa, as described in Appendix 3

#### 10.4.2. The solution of the boundary condition equation

The “boundary condition equation” can be written as follows, for the coupled 1 and coupled 2 model-families, respectively:

$$J_{n/2}(r_o\lambda_k) Y_{n/2}(r_1\lambda_k) - Y_{n/2}(r_o\lambda_k) J_{n/2}(r_1\lambda_k) = 0 \quad (105)$$

$$J_{n/2}(r_o\gamma_k) Y_{(n-2)/2}(r_1\gamma_k) - Y_{n/2}(r_o\gamma_k) J_{(n-2)/2}(r_1\gamma_k) = 0 \quad (106)$$

In the case of space dimension 1 ($n$=1) these are having the following roots for the coupled 1 and 2 model-families, respectively :

$$\lambda_k = \frac{k\pi}{(r_1 - r_0)} \quad (107)$$

$$\gamma_k = \frac{(2k-1)\pi}{2(r_1 - r_0)} \quad (108)$$

Inserting these into the analytical solution, the independent variables become non-dimensional. For $n$=3 or $n$=2 no closed form root formulae can be found. Solving the boundary condition equation and observing the numerical properties of the roots, two approximate formula can be found as follows ([41]).

**10.4.3. The approximate solution of the boundary condition equation**

The approximate formulae can analytically be derived by inserting the asymptotical Bessel function formulae into the boundary condition equation.The asymptotical Bessel functions formulae:

$$J_n(r) = \sqrt{\frac{2}{\pi r}} cos\left( r - \frac{\pi}{4} - \frac{n\pi}{2} \right) \quad (109)$$

$$Y_n(r) = \sqrt{\frac{2}{\pi r}} \sin\left( r - \frac{\pi}{4} - \frac{n\pi}{2} \right) \quad (110)$$

Inserting these into the boundary condition equation, the roots for the coupled 1 and 2 model-families, respectively:

$$\lambda_k \approx \frac{k\pi}{(r_1 - r_0)} \quad (111)$$

$$\gamma_k \approx \frac{(2k-1)\pi}{2(r_1 - r_0)} . \quad (112)$$

Inserting these roots into the analytical solution, two dimensionless arguments, a space variable (position factor) $R$ and a time variable (time factor) $T$, can be recognised in the foregoing approximate analytical solutions as follows for the coupled 1 and the coupled 2 models, respectively:

$$R = \frac{r}{r_1 - r_o} \text{ or } R = \frac{r}{2(r_1 - r_o)} \quad (113)$$

$$T = \frac{ct}{(r_1 - r_o)^2} \text{ or } T = \frac{ct}{4(r_1 - r_o)^2} \quad (114)$$

## 11. Appendix b Some consequences of the theoretical time factor

The coefficient of consolidation $c$ is determined by Teh and Houlsby (1988) as follows from their time factor:

$$c = \frac{T_{50}^{T-H}}{t_{50}} r_0^2 \; I_r^{1/2} \quad (115)$$

where $r_0$ is the radius of the CPT equipment, $t_{50}$ is the measured time for 50% dissipation, $I_r$ is the rigidity index. The time factor is heuristic, it is based on the observation that the theoretical dissipation curves can be normalized in terms of time.

A time factor can be derived from the analytical solution of the linear, coupled, cylindrical and spherical models in a mathematically precise way. The coupled 1 and 2 consolidation models imply the following time factors $T$, respectively (Imre et al., 2010):

$$T^1 = \frac{ct}{(r_1 - r_o)^2} \text{ or } T^2 = \frac{ct}{4(r_1 - r_o)^2} \quad (3a,b)$$

where $r_0$ is the radius of the rod, $r_1$ is the radius of the displacement domain. These are approximate in the cylindrical and spherical cases since the asymptotic Bessel formulae can only be used for this purpose. Compiling Equations (2) and (3a) for the coupled 1 model:

$$r_1 - r_0 = r_0 \left( T^{T-H} / T^1 \right)^{1/2} I_r^{1/4} \quad (4)$$

The approximate value of $r_1$ is 64.75 cm (according to the undrained strain path in Boston Blue Clay, see e.g. in Baligh, 1986, with 150 as rigidity index $I_r$) which is used in the evaluation in case of undrained penetration. The rigidity index $I_r$ measured in Szeged city for sandy silts is typically 200 to 800, for clays is typically 850 to 1000. It follows from eq (4) that $r_1$ - $r_0$ will be less for sands and silts than for clays by a factor of 0.7 to 0.95. It follows that some information can be collected for the value of $r_1$ in case of partly drained penetration using the rigidity index